\documentclass[fleqn,usenatbib]{mnras}

\usepackage[T1]{fontenc}

\DeclareRobustCommand{\VAN}[3]{#2}
\let\VANthebibliography\thebibliography
\def\thebibliography{\DeclareRobustCommand{\VAN}[3]{##3}\VANthebibliography}

\usepackage{graphicx}	
\usepackage{amsmath}	
\usepackage{amssymb}	
\usepackage{multirow} 
\usepackage{bm}

\newcommand{\UM}{\textsc{Universe}\textsc{Machine} }
\newcommand*\diff{\mathop{}\!\mathrm{d}}
\newcommand{\Msun}{\mathrm{M}_{\odot}}
\newcommand{\logMhalo}{\log(M_{\mathrm{vir}}/\Msun)}
\newcommand{\logMstellar}{\log(M_{\star}/\Msun)}
\newcommand{\tdelay}{t_{\mathrm{delay}}} 
\newcommand{\tperi}{t_{\mathrm{fp}}}

\newcommand{\tauenv}{\tau_{\mathrm{env}}}
\newcommand{\sbr}[1]{_{\textrm{#1}}}
\newcommand{\fQinfall}{f_{\textrm{Q,infall}}}

\newcommand{\deltaMWA}{ \delta\mathrm{MWA}_\mathrm{infall}}

\makeatletter
\DeclareRobustCommand{\HI}{%
  \mbox{H\,\check@mathfonts\fontsize\sf@size\z@\selectfont I }%
}
\makeatother

\usepackage{newtxtext,newtxmath}

\title[Constraining satellite quenching with stellar ages]{Constraining quenching timescales in galaxy clusters by forward-modelling stellar ages and quiescent fractions in projected phase space}

\author[A. M. M. Reeves et al.]{
Andrew M. M. Reeves$^{1,2}$\thanks{E-mail: andrew.reeves@uwaterloo.ca},
Michael J. Hudson$^{1,2,3}$,
Kyle A. Oman$^{4,5}$
\\
$^{1}$Department of Physics and Astronomy, University of Waterloo, Waterloo, ON N2L 3G1, Canada \\
$^{2}$Waterloo Centre for Astrophysics, University of Waterloo, Waterloo, ON N2L3G1, Canada \\
$^{3}$Perimeter Institute for Theoretical Physics, Waterloo, ON N2L 2Y5, Canada \\
$^{4}$Institute for Computational Cosmology, Durham University, South Road, Durham DH1 3LE, United Kingdom \\
$^{5}$Department of Physics, Durham University, South Road, Durham DH1 3LE, United Kingdom \\
}

\date{Accepted XXX. Received YYY; in original form ZZZ}

\pubyear{2022}

\begin{document}
\label{firstpage}
\pagerange{\pageref{firstpage}--\pageref{lastpage}}
\maketitle

\begin{abstract}
We forward-model mass-weighted stellar ages (MWAs) and quiescent fractions ($f_\mathrm{Q}$) in projected phase space (PPS), using data from the Sloan Digital Sky Survey, to jointly constrain an infall quenching model for galaxies in $\logMhalo > 14$ galaxy clusters at $z\sim 0$. We find the average deviation in MWA from the MWA-$M_\star$ relation depends on position in PPS, with a maximum difference between the inner cluster and infalling interloper galaxies of $\sim 1$~Gyr. Our model employs infall information from N-body simulations and stochastic star-formation histories from the \UM model.
We find total quenching times of $t_\mathrm{Q}=3.7\pm 0.4$~Gyr and $t_\mathrm{Q}=4.0\pm 0.2$~Gyr after first pericentre, for $9<\logMstellar<10$ and $10<\logMstellar<10.5$ galaxies, respectively. By using MWAs, we break the degeneracy in time of quenching onset and timescale of star formation rate (SFR) decline. We find that time of quenching onset relative to pericentre is $\tdelay=3.5^{+0.6}_{-0.9}$~Gyr and $\tdelay=-0.3^{+0.8}_{-1.0}$~Gyr for $9<\logMstellar<10$ and $10<\logMstellar<10.5$ galaxies, respectively, and exponential SFR suppression timescales are $\tauenv \leq 1.0$~Gyr for $9<\logMstellar<10$ galaxies and $\tauenv\sim 2.3$~Gyr for $10<\logMstellar<10.5$ galaxies. Stochastic star formation histories remove the need for rapid infall quenching to maintain the bimodality in the SFR of cluster galaxies; the depth of the green valley prefers quenching onsets close to first pericentre and a longer quenching envelope, in slight tension with the MWA-driven results. Taken together these results suggest that quenching begins close to, or just after pericentre, but the timescale for quenching to be fully complete is much longer and therefore ram-pressure stripping is not complete on first pericentric passage.
\end{abstract}





\section{Introduction}

Galaxies have long been known to exhibit a bimodality in their star formation rates (SFRs) such that the population of bluer galaxies with higher SFRs make up the `star-forming main sequence' and the redder galaxies with lower SFRs forming the quiescent `red sequence' \citep[e.g.][]{Bell2004,Brinchmann2004,Brammer2011,Muzzin2012}. The quiescent fraction increases strongly with increasing galaxy stellar mass, likely due to some internal (`mass') quenching process, as well as increasing with the density of its surrounding environment \citep[e.g.][]{Cooper2007,Peng2010}, due to some environmental quenching process(es). Despite significant exploration of quenching trends and their statistical relation to internal and environmental properties, it remains a matter of debate which quenching mechanisms are responsible for the bulk of quenching in the densest environments across cosmic time, particularly galaxy clusters \citep{Naab2017,Wechsler2018}.

Galaxies falling into groups or clusters may have their star formation affected by a number of physical processes. Initially, upon infall into a cluster, a galaxy may experience enhanced star formation, due to ram pressure compressing and triggering collapse of a galaxies' cold gas reservoir \citep[see e.g.][]{Jaffe2018,Vulcani2018jelly,Roberts2021}. If strong enough, ram pressure can strip the cold gas in star-forming galaxies, triggering relatively abrupt quenching \citep[normally $<1$~Gyr is assumed to be indicative of this scenario, see e.g.][]{Boselli2016, Fossati2018}. Alternatively, numerous studies support stripping of just the hot halo gas could be leading to a `strangulation' scenario, where a star-forming galaxy continues forming stars using its cold gas reservoir, but since its cold gas supply is not replenished, the galaxy then quenches once it runs out of cold gas \citep[such as argued by][]{Larson1980,Balogh2000,Bekki2002,Boselli2006,McGee2009,Taranu2014,Paccagnella2016}. Such a scenario is often simply parameterized by a `delayed-then-rapid' quenching prescription \citep{Wetzel2013}, with a range of time until quenching (after `infall' -- definitions for which vary) of $2-7$~Gyr preferred by observations at $z=0$ \citep[see also e.g.][among numerous others]{DeLucia2012,Moster2018EMERGE,Rhee2020,Oman2021}. 
Other effects, like harassment from galaxies passing near each other, as well as mergers, can affect the star formation in a galaxy \citep{Boselli2006}. Complicating this picture is the `pre-processing' of galaxies in host haloes of a lower mass groups prior to their infall into a cluster, which can additionally enhance the quiescent fraction in clusters compared to accreting isolated galaxies from the field. 
Although widely recognized as playing an important role in explaining the environmental dependence of star formation, work remains in characterizing the significance of pre-processing for quenching, where exactly it occurs, and how it evolves with redshift \citep{Fujita2004, McGee2009, DeLucia2012, Webb2020}. It is at least clear that quenching is significantly enhanced relative to the field at least up to $z\sim 1.5$ in rich galaxy groups \citep[][]{Gerke2007,Wetzel2012,Mok2013,Reeves2021} and clusters \citep{vanderBurg2018,PintosCastro2019,vanderBurg2020GOGREEN}.

Constraining timescales using forward modelling of observable quantities in projected phase space (PPS) in clusters can provide significant information for discerning when and where different physical quenching mechanisms may be dominant. In observational works, galaxy properties have shown a strong link to their position in PPS \citep[for a variety of distinct examples see e.g.][]{Mahajan2011,Smith2012ComaAges,Muzzin2014,Jaffe2015,Noble2016,Gavazzi2018,Jaffe2018,Kelkar2019,Kim2020,Rhee2020,Roberts2021}. Little detailed forward modelling work has been done on constraining quenching timescales in PPS, aside from that developed in the series of papers by Oman, Hudson and collaborators \citep{Oman2013orbitLibrary,Oman2016satQuenching,Oman2021}. The second and third of these papers use the quiescent fraction of cluster galaxies in PPS coordinates as well as corresponding distribution of infall times inferred from N-body simulations. The PPS coordinates used in this context, namely the radial distance from cluster center and line-of-sight velocity offset from its host system, only carry magnitude (i.e. no sign) information in practice, as it's generally not possible to distinguish between a satellite receding towards a background host or a background satellite receding away from a foreground host. This method provides a nearby interloper population on the cluster outskirts (outside some 3D radial cut) against which to compare cluster satellites. This has an advantage over using a generic field population: using these interlopers in the cluster outskirts provides an already pre-processed infalling galaxy sample, allowing better isolation of cluster physics from the physics involved in pre-processing. 

Whereas many previous works that modeled environmental quenching have relied on quiescent fractions, observational indicators that are sensitive to the star formation history over longer timescales should provide more information. Luminosity- or mass-weighted ages (MWAs) in particular not only provide complementary quantitative information about the star formation history of galaxies, but are also insensitive to whether a given galaxy's star formation history is assumed to be smooth or stochastic (see discussion in Section~\ref{sec:ssfr-bimodality-stochasticity}).

Studies using spectroscopic indicators of stellar age have long established that more massive galaxies, or, more accurately, galaxies with deeper potential wells, have older ages \citep{NelanSmithHudson2005, GravesFaberSchiavon2007}. Dependence on environment is weaker: early studies focused on the age difference between field and cluster ellipticals after controlling for mass or velocity dispersion \citep{ThomasMarastonBender2005,BernardiNicholSheth2006}. Radial trends of the stellar ages of galaxies within clusters (at fixed velocity dispersion) were found \citep{SmithHudsonLucey2006} with a stronger environmental dependence on the ages of dwarf galaxies \citep{Smith2012ComaAges}.

Studies of stellar age in PPS are less common. \citet{Pasquali2019} considered luminosity-weighted stellar ages in zones of PPS around low redshift clusters, finding ages that are older by $\sim 2$~Gyr for the innermost lower-velocity regions of PPS compared to the outermost regions. Recently, \cite{KhullarBaylissGladders2022} studied D$4000$-derived ages in PPS in $0.3<z<1.1$ clusters, again finding older ages for PPS regions associated with the core of the cluster \citep[see also][]{Kim2022}.

There has been even less effort modelling the effect of quenching on the stellar ages. \citet{Taranu2014} used infalling subhalo orbits from $N$-body simulations to forward-model the age-sensitive Balmer line Lick indices as a function of cluster-centric radius. Their models preferred long quenching timescales occurring for galaxies that have passed within $\sim 0.5 r_{200c}$ (i.e. preferring slow `strangulation' over rapid ram-pressure stripping).

As little work has been done forward modelling spectroscopically-derived ages in clusters and none has been carried out in full PPS with a large sample \citep[but see][for a first attempt with a sample of 11 galaxies in the Coma cluster]{Upadhyay2021}, the goal of this work is to extend the modelling of \citet{Oman2016satQuenching} and \citet{Oman2021}, who studied the quiescent fraction, $f\sbr{Q}$, in PPS. \citet{Oman2021} found that using $f\sbr{Q}$ alone is not sufficient to constrain both the time of quenching onset and the timescale over which quenching occurs on average and so they marginalized over this latter parameter to find the average time at which 50~per~cent of infalling star-forming galaxies have quenched. Extending this and similar work could make use of, for example, the entire sSFR distribution, but as we will describe, the shape of the quiescent bump is sensitive to observational systematics, while the depth of the green valley is sensitive to choice of star formation history (smooth versus stochastic). In this paper, we will focus on extending \citet{Oman2021} by forward modelling the observed MWAs of galaxies in PPS, in addition to using quiescent fractions. This additional observable will enable joint constraints on both the time of onset of infall quenching and the timescale for the duration of quenching preferred by both observables, for the population of galaxies as a whole.

Our paper is structured as follows. In Section~\ref{sec:data-and-sims}, we describe how we parameterise PPS and the SDSS-based datasets, from which we get our $f\sbr{Q}$ and mean MWAs. We then, in  Section~\ref{sec:modelling-and-results}, introduce our forward modelling approach (including sources of star formation histories and infall histories) and present our quiescent fraction and mean MWA modelling results in PPS as well as the constraints on quenching timescales they provide. In Section~\ref{sec:discussion}, we discuss the robustness of our results, how they compare to literature, and what they imply for the dominant infall quenching processes. We then conclude in Section~\ref{sec:conclusions}.

Unless otherwise specified, the following assumptions and conventions are used. Uncertainties are estimated from the 16\textsuperscript{th}-84\textsuperscript{th} percentile interval (equivalent to 1-$\sigma$ for a Gaussian-distributed variable). Logarithms with base 10 ($\log_{10}$) are written simply as `$\log$' throughout this work. A flat $\Lambda$CDM cosmology consistent with the Planck 2015 cosmological parameters \citep{Planck2015CosmoParams} is assumed, namely $H_0=68~{\rm km}~{\rm s}^{-1}~{\rm Mpc}^{-1}$, $\Omega_m=0.31$, and $\Omega_{\Lambda}=0.69$. We define a virial overdensity at $z=0$ in this work as $\sim 360$ times the background density, $\Omega_m \rho_c$, corresponding to the density of a recently virialized spherical top-hat solution \citep[see e.g.][]{BryanNorman1998}
\footnote{For a commonly used virial overdensity definition of $200\rho_c$, this corresponds to $M_{200c}/M_{\mathrm{vir}}\approx 0.81$ and $r_{200c}/r_{\mathrm{vir}}\approx 0.73$ \citep{Oman2021}, assuming a concentration parameter of $c_{200}=r_{200c}/r_s\approx 5$ (where $r_s$ is the NFW profile scale radius), a typical value for clusters \citep{Ludlow2014}.}. A \citet{Chabrier2003} initial mass function (IMF) is assumed throughout.


\section{Data and Simulations}\label{sec:data-and-sims}

\subsection{PPS selection and conventions} \label{sec:PPS-selection}

We adopt the convention of \citet{Oman2013orbitLibrary} for selecting objects in PPS to allow direct comparison to observations: there are two sets of cluster-centred coordinates: $(r, v)$ for the $2\times 3\mathrm{D}$ physical phase space coordinates and $(R, V)$ for the normalized PPS coordinates. Specifically, the PPS coordinates consist of the distance to cluster centre perpendicular to the line-of-sight and the velocity component along the line of sight, which we choose to be the Cartesian $Z$-axis of a given simulation dataset. The projected radius between the cluster centre and a galaxy is then
\begin{equation}
    R = \bigg(\frac{1}{r_{\mathrm{vir,\mathrm{3D}}}}\bigg) \sqrt{(r_X - r_{X,\mathrm{cls}})^2 + (r_Y - r_{Y,\mathrm{cls}})^2}
\end{equation}
and the projected velocity relative to the motion of the cluster centre of mass is 
\begin{equation}
    V = \bigg( \frac{1}{\sigma_{\mathrm{3D}}} \bigg) \left|(v_Z - v_{Z,\mathrm{cls}}) + H(Z - Z_{\mathrm{cls}})\right| .
\end{equation}
Using an absolute value for the projected velocity definition captures an assumption that the distances to clusters and satellites are not measured with enough accuracy to determine the sign of their relative velocity. The additional $H$ term corrects for the Hubble flow. To allow easy comparison between clusters, for both simulations and observations, we have normalized the PPS coordinates. For ease of comparison to \citet{Oman2016satQuenching} and \citet{Oman2021}, radial coordinates are normalized by dividing by the 3D virial radius of the cluster, $r_{\mathrm{vir,\mathrm{3D}}}$, and velocity coordinates by the 3D velocity dispersion, $\sigma_{\mathrm{3D}}$, of a given cluster. 
As we are averaging over many clusters in both the orbit library and observed sample, we can assume that clusters are approximately spherically symmetric. This in turn allows relation of the 3D velocity dispersion to the observable 1D velocity dispersion by $\sigma_{\mathrm{3D}} \approx \sigma_{\mathrm{1D}}/\sqrt{3}$. We note that in order to ensure consistency between the normalization factors used for $R$ and $V$, they should not be independent, i.e. the velocity dispersion should be derived from the virial radius or vice-versa. See Section~\ref{sec:SDSS_dataset_description} for details of how these are calculated for the SDSS data.

\subsection{Infall histories: N-body simulation orbit library} \label{sec:oman-orbit-library}

\begin{figure}
	\includegraphics[width=\columnwidth]{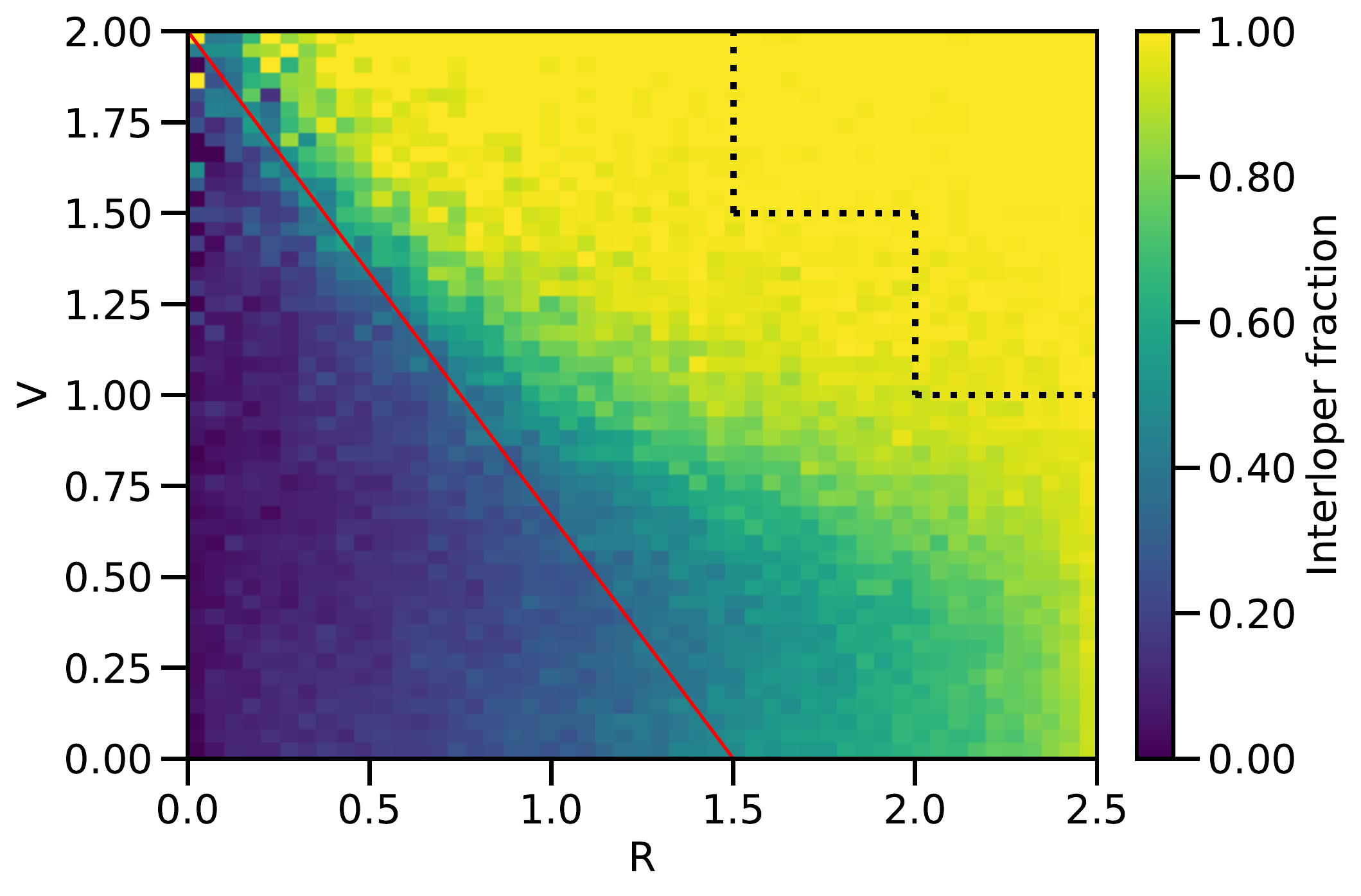}
    \caption{
        The fraction of interlopers in (R,V) projected phase space bins determined using the orbital library from \citet{Oman2021}. $R$ is distance from cluster centre, in units of $r_{\mathrm{vir,\mathrm{3D}}}$ and $V$ is the velocity offset from cluster centre, with units of $\sigma_{3\mathrm{D}}$. The red diagonal line, $V=-(4/3)R+2$, guides the eye as to where $\sim 50$~per~cent of galaxies are interlopers, aside from a portion around $(R,V)\sim (1.2,0.4)$, which has many galaxies at or close to their first apocentre after entering the cluster. The dotted line delineates a region in the upper right that is heavily dominated by the interloper population, which we use to compare between SDSS and \UM interlopers.
    }
    \label{fig:interloper_fraction_PPS_distribution}
\end{figure}

We use the PPS orbital libraries from \citet{Oman2021}; they base their methodology on \citet{Oman2016satQuenching} to produce a catalogue of orbit parameter probability distributions from an N-body simulation. As described in \citet{Oman2021}, the `level 0' voids-in-void-in-voids simulation \citep[VVV;][]{Wang2020VVV} is used, which has a box size of $500 ~h^{-1} \mathrm{Mpc}$ on a side, mass resolution elements of $10^9~h^{-1} \Msun$, and a force softening scale of $4.6~h^{-1}\mathrm{kpc}$. They run the simulation to scale factor $a=2$, corresponding to redshift $z = -0.5$ or $\sim 10~\mathrm{Gyr}$ into the future (i.e. negative lookback times), allowing them to find the time of first pericentre for $>99.9$~per~cent of resolved satellite galaxies. Host halo masses and satellite halo masses are estimated using the \textsc{rockstar} halo finder; infall times and times of first pericentre are derived from halo merger trees generated by the \textsc{Consistent Trees} utility. These merger trees include all haloes with $>30$ particles, corresponding to $M_{\mathrm{vir}}\gtrsim 4\times 10^{10}~\Msun$, allowing resolution of $M_{\mathrm{vir}}\sim 2.5\times 10^{11}~\Msun$ haloes, which host the lowest stellar mass galaxies in our observed sample -- $M_{\star}\sim 10^{9.5}~\Msun$ (at least until they are stripped of $\gtrsim 85$~per~cent of their halo mass). 

Satellites are identified as haloes within a 3D aperture of $2.5~r_{\mathrm{vir}}$ at $z=0$ for $\log (M_{\mathrm{vir}}/\Msun)>12$ host systems (although we only use hosts with $\log (M_{\mathrm{vir}}/\Msun)>14$ in this work). Satellite primary progenitors/descendants are traced backward/forward through time and orbits relative to the $z=0$ primary progenitor/descendent of their host are recorded. The time of first pericentre is not interpolated (it is very challenging to do this properly and is unnecessary for our purposes), so it is worth noting here that this results in non-uniform output times, with a median timestep of $220~\mathrm{Myr}$ and a maximum timestep of $380~\mathrm{Myr}$, sufficient to resolve the characteristic quenching and stripping timescales fit in \citet{Oman2021}. Worth noting is that the satellite halo mass, $M_{\mathrm{sat}}$, is its maximum mass at $z\geq 0$, since maximum halo mass is better correlated with stellar mass in `moderately stripped' satellites; see e.g. \citet{Conroy2006} and Appendix~A of \citet{Wetzel2013} for further elaboration on this.

For the orbit library itself, since satellites are all galaxies within $2.5 r_{\mathrm{vir}}$, interlopers are then naturally defined as all galaxies that fall within $2.5 r_{\mathrm{vir}}$ in projection, but outside $2.5 r_{\mathrm{vir}}$ in 3D. Only a vanishingly small fraction, $\ll 1$~per~cent, of interlopers would have been classified as satellites at an earlier time. All satellites have a recorded time of first infall into the final cluster ($2.5 r_{\mathrm{vir}}$ in 3D) and time of first pericentre ($\tperi$) in the final cluster; we only use $\tperi$ in our modelling and analysis. We make use of their relative abundance compared to selected satellites to define the probability that an `observed' galaxy is an interloper. For the sake of illustration and intuition for the reader, we plot the statistical fraction of galaxies that are interlopers at a given position in PPS, rather than cluster satellites, in Fig.~\ref{fig:interloper_fraction_PPS_distribution}. In the plot it's clear that at both high $R$ and $V$ the vast majority of galaxies are interlopers and vice-versa for lower $R$ and $V$. We note that around $(R,V)=(1.5-2,0-0.5)$ there are many galaxies on their first infall into a cluster, whereas galaxies at around $(R,V)=(0-0.5,1-2)$ are primarily galaxies at their first pericentre. Over several orbits galaxies settle to low $(R,V)$. As well, we plot a straight line indicating where approximately 50~per~cent of galaxies are interlopers -- this should provide a helpful rule-of-thumb for the reader in the plots of various quantities in PPS that will follow.

\subsection{Star formation histories: the \UM} \label{sec:UM-SFHs}

For individual galaxies' SFR histories we use the \UM semi-analytic model \citep{Behroozi2019UniverseMachine} 
which
parametrises galaxy SFRs as a function of halo potential well depth, redshift, and assembly history. Their halo potentials and assembly history are derived from the \textit{Bolshoi-Planck} dark matter simulation \citep{Klypin2016,RodriguezPuebla2016}, which features a periodic comoving volume $250h^{-1}$~Mpc on a side with $2048^3$ ($\sim 8\times 10^9$) particles. The simulation employs a flat $\Lambda \mathrm{CDM}$ cosmology compatible with Planck15 results \citep{Planck2015CosmoParams}, with stored snapshots equally spaced in $\log(a)$ (180 intervals). The \textsc{rockstar} halo finder \citep{Behroozi2013ROCKSTAR} and \textsc{Consistent Trees} \citep{Behroozi2013CONSISTENTTREES} are used to construct merger trees, the same codes used in the construction of the merger trees for the orbit library employed in this work (Section~\ref{sec:oman-orbit-library}).

For our purposes, it was important that their model reliably reproduce the observed SFRs and quiescent fractions over time, especially in overdense regions in SDSS (although not necessarily cluster cores, as we are only using \UM interloper galaxies, described in the following paragraph).
A key aspect of the \UM model is that SFRs are stochastic -- they fluctuate over time -- and are linked to the merger history of the halo. \citet{Behroozi2019UniverseMachine} found that without such scatter the quiescent fractions of satellites would be too high (relative to centrals). Specifically, higher SFRs are assigned to halos with higher levels of halo growth, parameterised by the relative logarithmic growth in $v_\mathrm{max}$, over the past dynamical time (where $v_\mathrm{max}$ is the maximum circular velocity of the halo).  
The stochasticity has two components: a short timescale component ($\sim 10-100$~Myr) representing internal processes affecting local galactic cold gas, and the second component linked to the dynamical time ($\sim 1$~Gyr).
This stochasticity will be a key difference in our comparison to other infall models (which use smooth star formation histories).
A galaxy can transition between quiescent and star-forming, as long as the overall quiescent fraction and other properties are matched to the various observations they used, in a self-consistent manner.

We construct a PPS catalogue using the \UM simulation results at $z=0$, assuming arbitrarily that the $z$-axis is the observational line-of-sight. For each of the 144 \UM simulation sub-volumes, galaxies were selected from around central haloes (`\texttt{UPID}'==-1) with halo masses $\logMhalo > 14$. All galaxies (including centrals) are selected so that their projected radius is $R<2.5$ and projected velocity is $V < 2$, where the PPS coordinates $(R, V)$ are defined as described previously in Section~\ref{sec:PPS-selection}. True interlopers and satellites, as opposed to interlopers and satellites as defined in PPS, are identified with 3D cuts of $R>2.5$ and $R<2.5$, respectively.

\subsection{SDSS dataset} \label{sec:SDSS_dataset_description}

For the observed sample of galaxies, we build on the sample used by \citet{Oman2021} for their quenching analysis. As in \citet{Oman2021}, we use the SDSS Data Release 7 catalogue \citep{Abazajian2009}, supplemented with SFRs \citep{Brinchmann2004,Salim2007} and stellar masses \citep{Mendel2014}, which were used in \citet{Oman2021} for estimating galaxy subhalo masses using subhalo abundance matching. We also use the spectroscopic sample of galaxies, which we note suffers from fiber collisions, although \citet{Oman2021} determine through a detailed exploration that the impact of this should be minimal.

We use the mass-weighted ages, as well as the accompanying stellar mass estimates, of \citet{Comparat2017}, who performed full spectral fitting of galaxy properties using \textsc{FIREFLY} \citep{Wilkinson2017FIREFLY}. We compare their stellar masses with those of \citet{Mendel2014} and confirm that offsets in the stellar masses do not affect our results. In particular, we use the fits done using the stellar population models of \citet{MarastonStromback2011}, a \citet{Chabrier2003} IMF, and the \textsc{MILES} stellar library \citep{SanchezBlazquez2006,FalconBarroso2011,Beifiori2011} for the mass-weighted ages (`CHABRIER\_MILES\_age\_massW'), and stellar masses (`CHABRIER\_MILES\_stellar\_mass').

For simplicity, we focus on high-mass cluster haloes, $\logMhalo > 14$, where environmental effects should be most extreme. As in \citet{Oman2021}, we select candidate satellite galaxies around groups and clusters from the \citet{Lim2017} and \citet{vonderLinden2007} catalogues, with cluster virial masses peaking at $\sim 3 \times 10^{13} \Msun$ and $\sim 3\times 10^{14}\Msun$, respectively. We exclude groups with redshifts $z<0.01$, as their members are generally too bright to be covered by SDSS spectroscopy. We note that the two group catalogues are constructed with different algorithms. \citet{vonderLinden2007} look for overdensities of galaxies with similar colours, while \citet{Lim2017} use a friends-of-friends group finding algorithm. This does not significantly affect our analysis as we only use the cluster masses, centres, and line of sight velocities from the catalogues, rather than membership information. Satellite velocity offsets are normalized using cluster velocity dispersions, calculated from the virial masses of the clusters in the catalogues. The velocity dispersions are similar to the dark matter particle velocity dispersion of the systems \citep[within $\lesssim 10$~per~cent; see][table 1]{Munari2013}; this is important since the velocity dispersion of dark matter particles in the host halo is used to normalize the velocity offsets of satellite haloes in the simulations that the \citet{Oman2021} orbit library is derived from (see Section~\ref{sec:oman-orbit-library}). Host halo masses are calculated following eq.~1 of \citet{vonderLinden2007} and eq.~4 of \citet{Lim2017}. These are converted to virial masses assuming an NFW density profile \citep{navarro1997universal} and mean mass-concentration relation from \citet{Ludlow2014}, with differences between assumed cosmologies accounted for. The mean enclosed density is then used to define the virial radii,
\begin{equation}
    r_{\mathrm{vir}} = \bigg( \frac{3}{4\pi} \frac{M_{\mathrm{vir}}}{\Delta_{\mathrm{vir}}(z) \Omega_m(z) \rho_{\mathrm{crit}}(z) } \bigg)^{\frac{1}{3}} ,
\end{equation}
where $\Delta_{\mathrm{vir}}(z)$ is the virial overdensity in terms of the mean matter density $\Omega_m(z)\rho_{\mathrm{crit}}(z)$, with critical density $\rho_{\mathrm{crit}} = 3H^2 / (8\pi G)$. Group velocity dispersions are defined following \citet{Biviano2006}, but with redshift dependence from \citet{BryanNorman1998}; explicitly this is
\begin{equation}
    \frac{\sigma_{\mathrm{1D}}}{\mathrm{km~s}^{-1}} = \frac{0.0165}{\sqrt{3}} \bigg( \frac{M_{\mathrm{vir}}}{\Msun} \bigg)^{\frac{1}{3}} \bigg( \frac{\Delta_{\mathrm{vir}}(z)}{\Delta_{\mathrm{vir}}(0)} \bigg)^{\frac{1}{6}} (1+z)^{\frac{1}{2}} .
\end{equation}

We use two stellar mass bins, $9<\logMstellar<10$ and $10<\logMstellar<10.5$. We note that for $\logMstellar>10.5$ there are too few star forming galaxies to yield robust results, so we do not consider this mass range in this paper.

\subsubsection{Quiescent fraction trends in PPS around SDSS clusters}\label{sec:fQ-SDSS}

\begin{figure*}
	\includegraphics[width=\columnwidth]{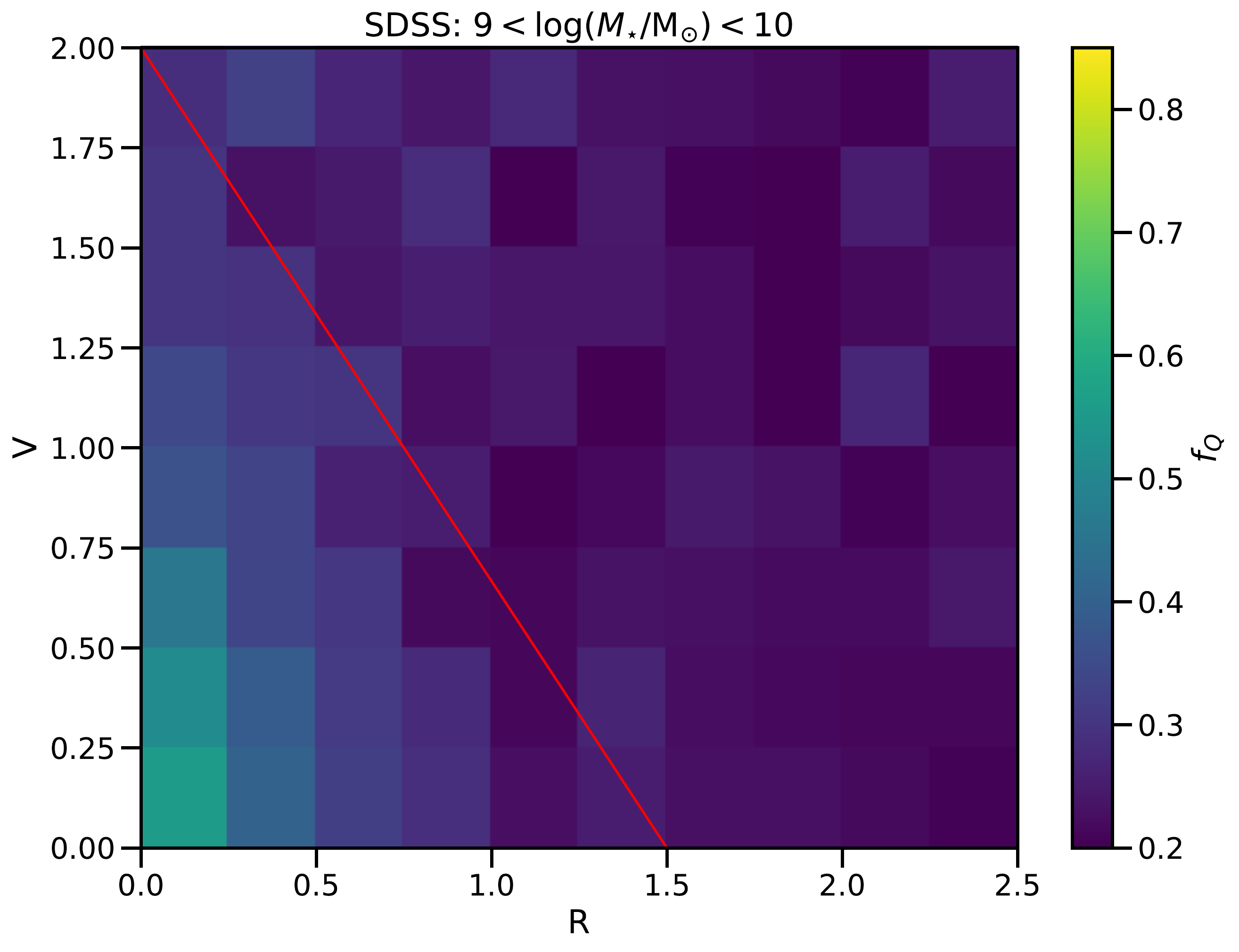}
	\includegraphics[width=\columnwidth]{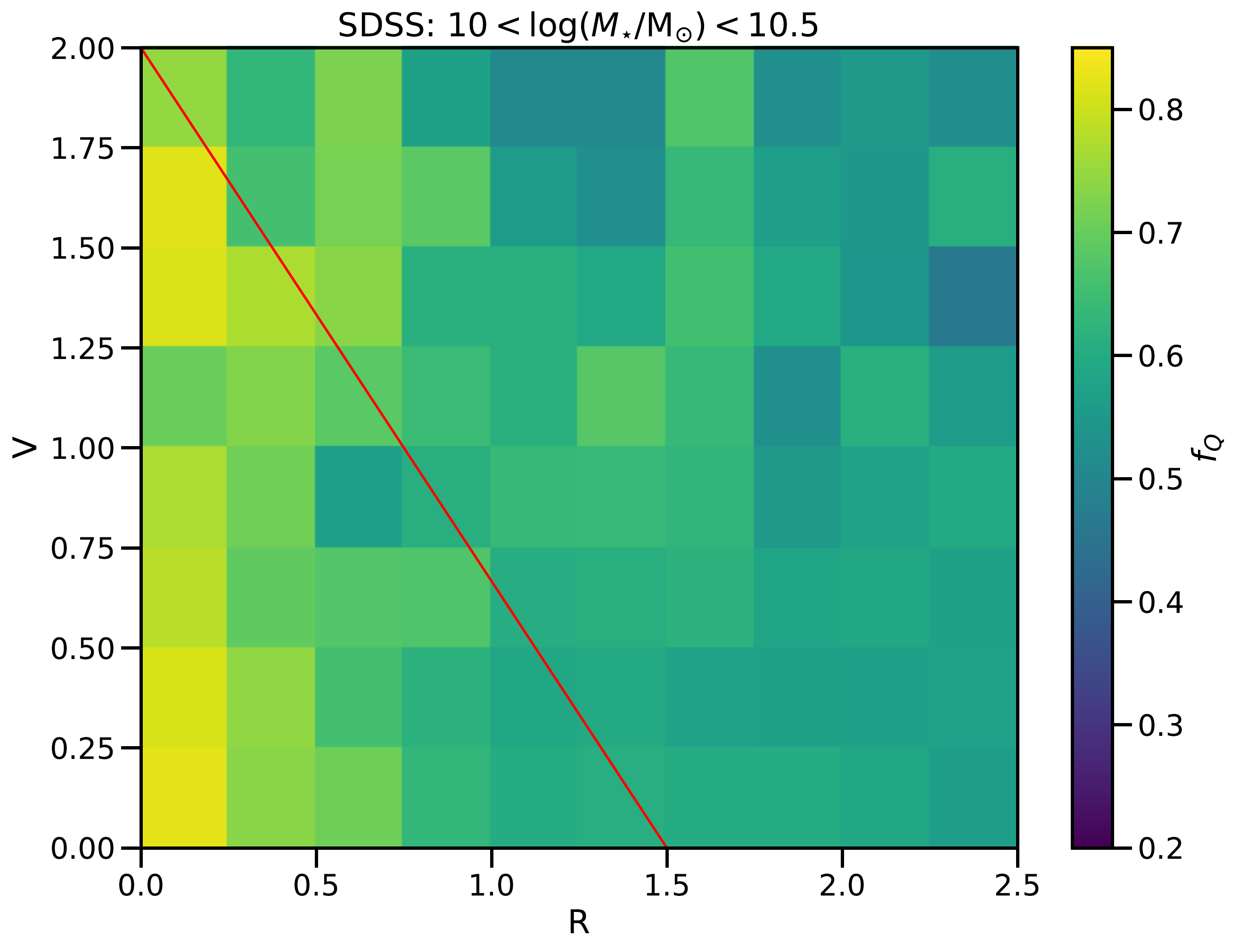}
    \caption{Quiescent fraction of SDSS galaxies in the stellar mass ranges $9<\logMstellar<10$ (left) and $10<\logMstellar<10.5$ (right) in clusters with host halo mass $\logMhalo>14$, binned in PPS coordinates. The quiescent fraction is significantly higher in cluster centres than in their outskirts ($\gtrsim 30$ per cent). For the lower stellar mass bin, regions within the $\sim 50$~per~cent interloper (red diagonal) line, show at least modestly elevated quiescent fractions, with little evidence of enhanced quenching immediately outside of (above) this line. For the higher stellar mass bin, some higher velocity galaxies within the virial radius also show somewhat elevated quiescent fractions.}
    \label{fig:fQ_SDSS}
\end{figure*}

Our first source of information for deriving quenching timescales is the quiescent fraction in SDSS clusters. As we are using the same observational data sample as \citet{Oman2016satQuenching} and \citet{Oman2021}, we define define star-forming and passive galaxies using the same cut, namely
\begin{align}
    \log(\mathrm{sSFR/yr}^{-1}) = -0.4 \log (M_{\star}/\Msun) - 6.6 ,
\end{align}
with star-forming (quiescent) galaxies lying above (below) the specific star formation rate (sSFR) relation.

We illustrate the quiescent fraction trends in PPS for $9<\logMstellar<10$ galaxies in Fig.~\ref{fig:fQ_SDSS}. In the figure we see well-known observed trends of decreasing $f\sbr{Q}$ with increasing distance from cluster centre as well as higher $f\sbr{Q}$ for galaxies with lower velocity offsets. We draw a line indicating roughly where $\sim 50$~per~cent of galaxies are interlopers, with galaxies above this line increasingly dominated by interlopers, moving from lower to higher R and V coordinates. Enhanced quenching largely falls within the region bounded by this 50~per~cent interloper line in PPS. Contrasting with the field results of \citet[][see upper right panel of e.g. fig.~10 to see star-forming fraction versus stellar mass for the SDSS field]{Oman2021}, with field defined as the average overall SDSS population, we see that the interloper quiescent fraction is higher than the field by $0.05$ and $0.09$ for our lower and higher stellar mass bins, respectively.

Having information on how quenching depends on both the radial position and velocity relative to host cluster centre gives us very useful information on when and where quenching occurs. It's clear in Fig.~\ref{fig:fQ_SDSS} that most quenching of infalling star forming galaxies occurs well within the virial radius. Modestly enhanced quenching up to high velocities indicates quenching is at least beginning to occur within the first infall or just after first pericentre. Although we have some useful information from $f\sbr{Q}$ in PPS, this information alone results in significant degeneracies between time of quenching onset and duration of quenching, as we will clearly show later in this work (see Section~\ref{sec:fQ-modelling-results}). To break this timescale information degeneracy we must turn to another independent observable as a function of its position in PPS.

\subsubsection{Mass-weighted age -- stellar mass relation} \label{sec:MWA_residual_definition}\label{sec:ages-data}

\begin{figure*}
	\begin{tabular}{cc}
      SDSS & \UM \\
      \includegraphics[width=\columnwidth]{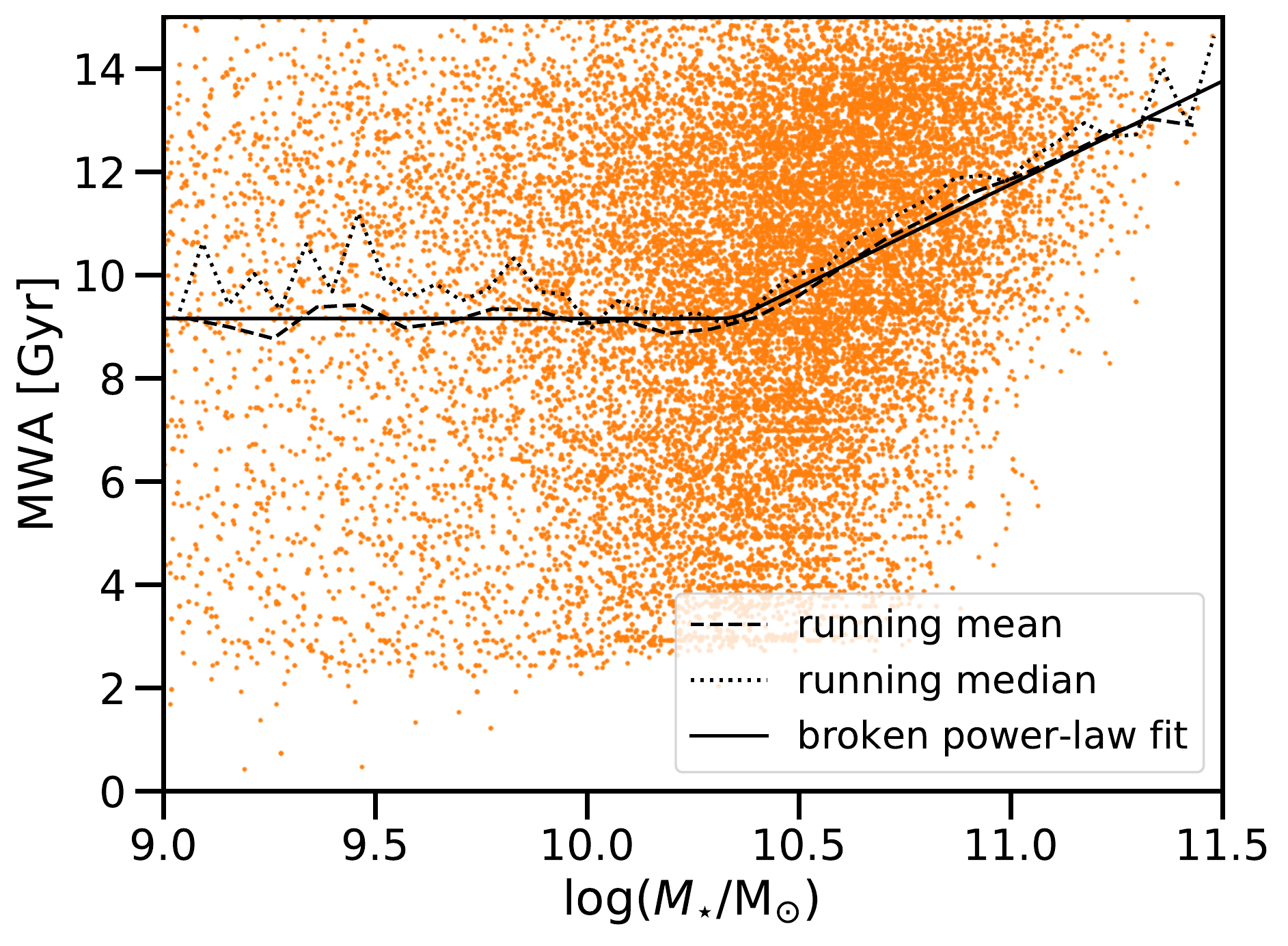} &  
      \includegraphics[width=\columnwidth]{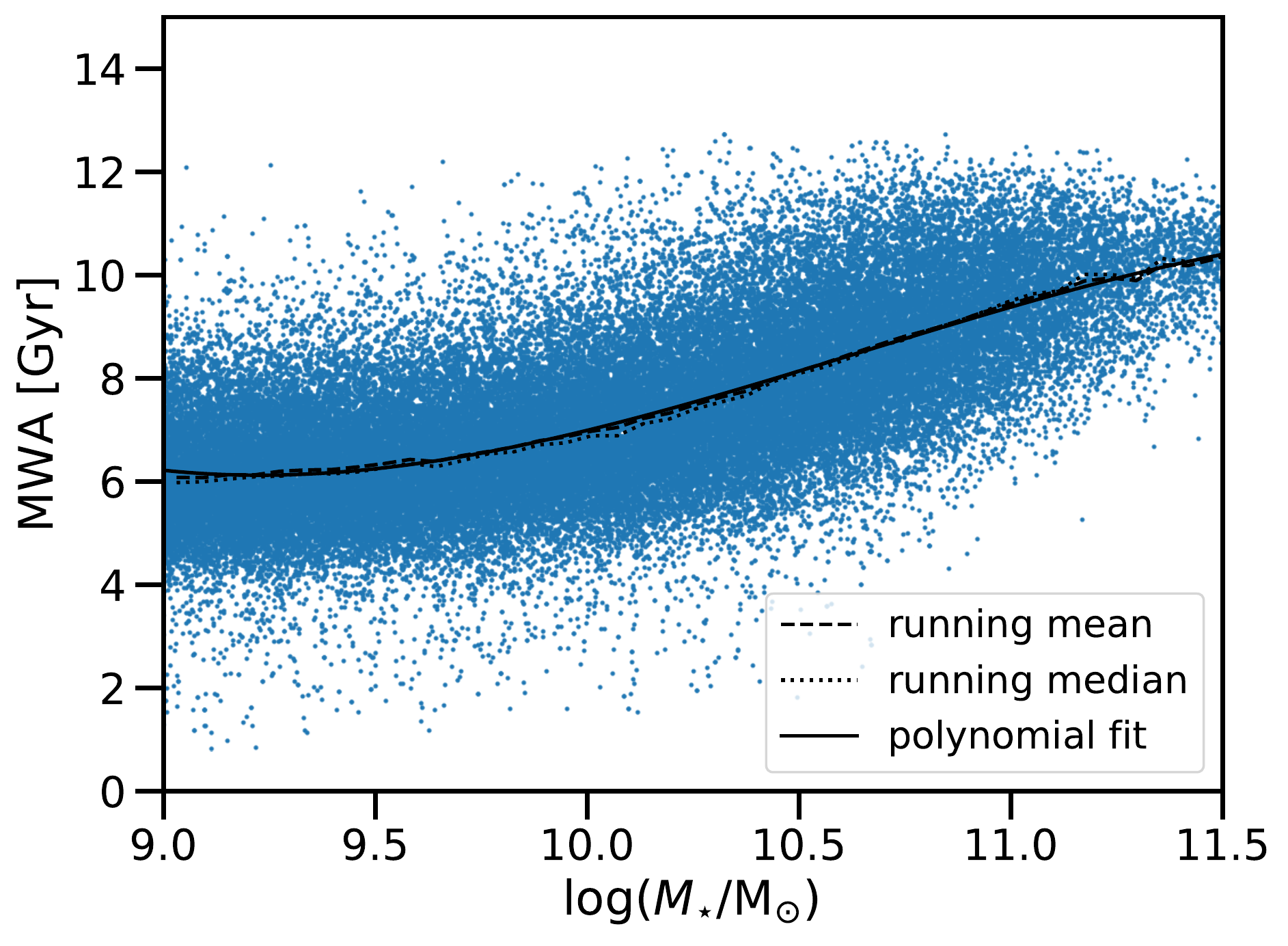} \\
     \includegraphics[width=\columnwidth]{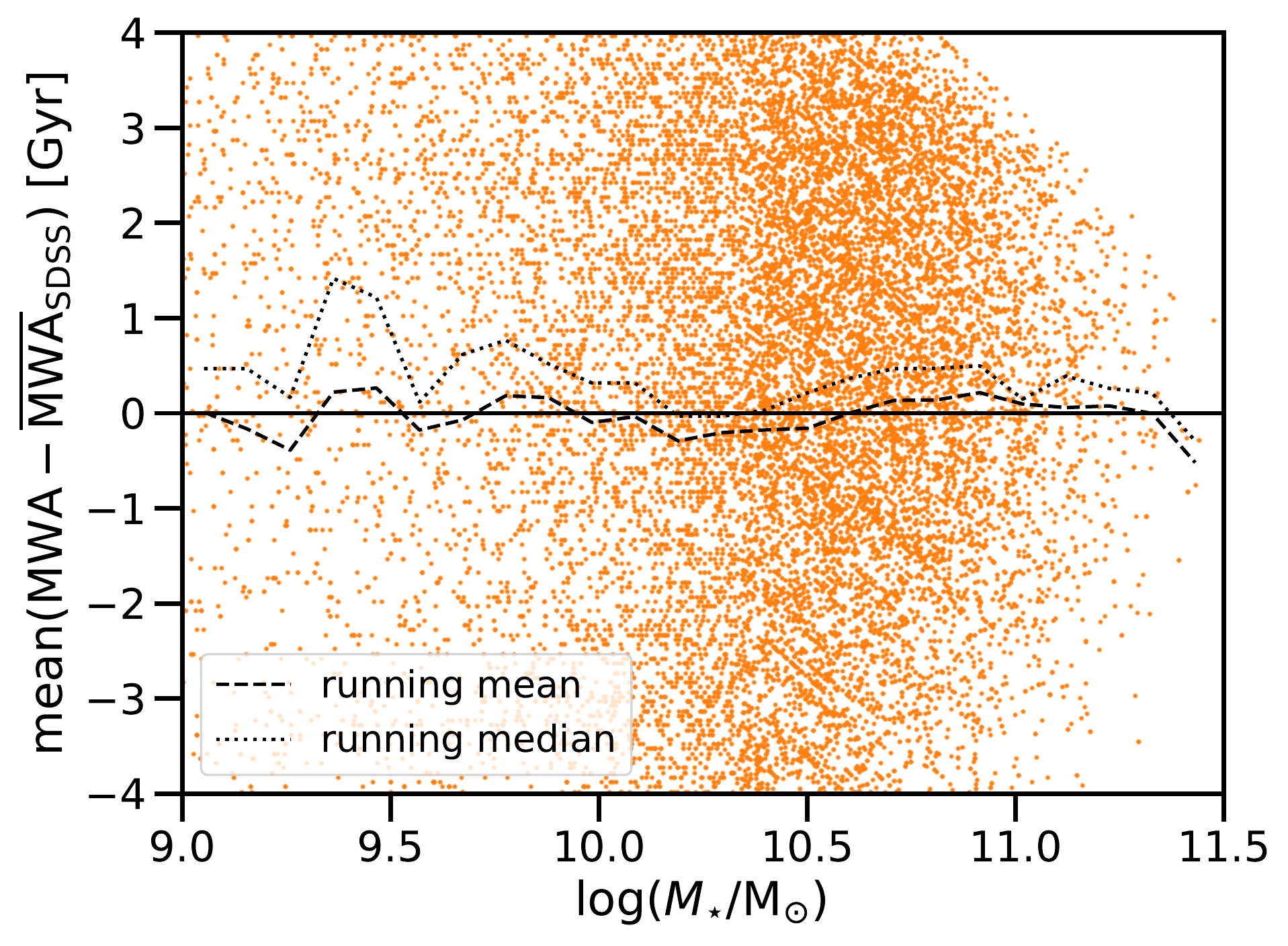} &   \includegraphics[width=\columnwidth]{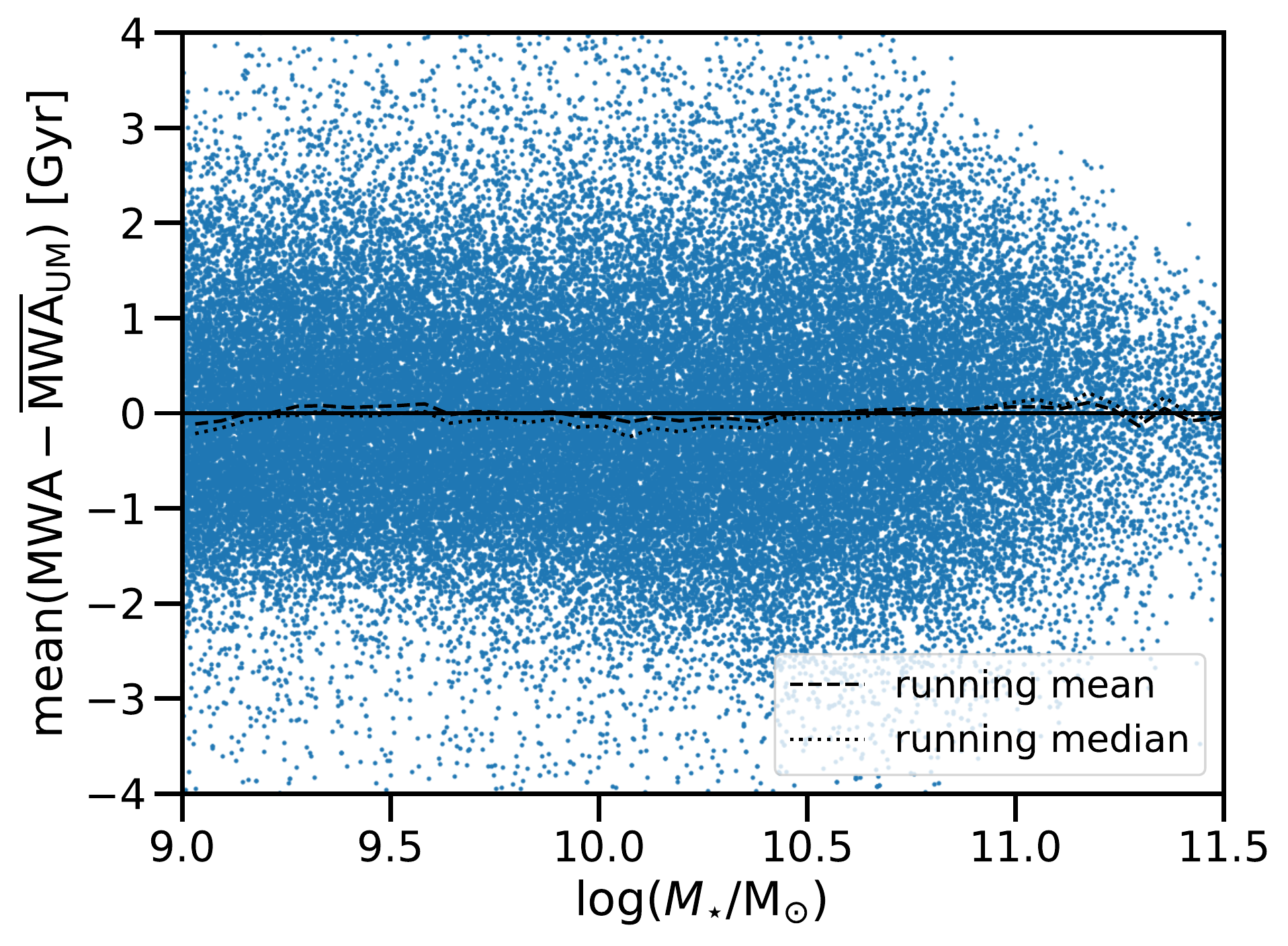} \\
    \end{tabular}
    \caption{\textit{Top panels:} SDSS observationally-derived mass-weighted ages for quiescent galaxies (left) exhibit a high degree of scatter and trend older with increasing galaxy stellar mass. \UM simulated galaxies (right) show a similar trend in mass-weighted age with stellar mass, albeit with a $\sim$ 3 Gyr offset and significantly less scatter (see Section~\ref{sec:ages-data} for discussion). SDSS points indicate the best-fitting mass-weighted ages for SDSS galaxies from \citet[][stellar masses are from \citealp{Comparat2017}]{Wilkinson2017FIREFLY}, with the running mean (median) shown as a dashed (dotted) line. The functional fits of the mean MWA-$M_{\star}$ relation used in the analysis of this work is fit to the running mean mass-weighted age at a given stellar mass and is shown with the solid black curves. \textit{Bottom panels:} Residual MWA from the respective MWA-M$_\star$ functional fits for SDSS (left) and \UM (right) is shown along with the running mean (dashed) and median (dotted) of the residual.}
    \label{fig:MWA_Mstellar}
\end{figure*}

For the stellar ages of SDSS galaxies we use a simple proxy to capture when the bulk of the stellar mass in a galaxy was formed, namely the mass-weighted age (MWA). We include only quiescent galaxies for our MWA-based measurements, since the emission lines of star-forming galaxies fill in the age-sensitive Balmer absorption lines, making them rather unreliable as an age indicator. In this subsection, we note a trend in MWA with stellar mass in both the observationally-derived SDSS values and the \UM simulated galaxies. As we are only interested in the differential change in MWA relative to an infalling population, we first describe how we control for the MWA-$M_{\star}$ effect in this work. After this, we will briefly illustrate observed MWA trends in PPS in Section~\ref{sec:data-MWA-in-PPS}.

We show the trend in MWA with stellar mass for both the observationally-derived SDSS values and the \UM \citep{Behroozi2019UniverseMachine} simulated galaxies side-by-side in Fig.~\ref{fig:MWA_Mstellar}. The mean and median trends that we see for SDSS and \UM are qualitatively similar, but the \UM galaxies are offset in overall MWA such that they are younger than SDSS MWAs by $\sim 3$ Gyr on average. We note that mass-weighted ages derived from spectra depend on the assumed star formation history. For example, \cite{AllansonHudsonSmith2009} showed that a range of star formation histories can fit the observed stellar absorption lines equally well in quiescent cluster galaxies, yielding mass-weighted ages that range from 5 Gyr (single burst) to 8 Gyr (exponentially declining star formation history) to 12.8 Gyr (`frosting' model) for the lowest-mass cluster galaxies. Since we use ages only in a differential sense in our analysis, i.e.\ comparing the difference in MWA in the cluster core to the infalling regions in PPS, any systematic offset in the ages due to different star-formation history assumptions will not affect our analysis. For SDSS, the MWA--$M_{\star}$ relation is approximately flat, with MWA $\sim$ 9 Gyr for $9<\logMstellar \lesssim 10.5$ galaxies, which then increases for $\logMstellar \gtrsim 10.5$ to $\sim$ 12-13 Gyr for the highest stellar mass galaxies, $\logMstellar > 11$. The increase in MWA for \UM galaxies from lowest to highest stellar masses shown in Fig.~\ref{fig:MWA_Mstellar} is similar in magnitude to that for SDSS ($\sim 4$ Gyr), albeit with a more steadily increasing monotonic trend, i.e. a non-flat trend in MWA below $\logMstellar < 10.5$. The scatter is also quite noticeably different -- the SDSS scatter can be thought of as the natural distribution of galaxies (shown modeled here by the \UM) convolved with the scatter induced in estimated MWAs from uncertainties in SED-fitting the galaxy spectra.

To remove the systematic trend in MWA with stellar mass and the $\sim 3$~Gyr offset between the SDSS observationally-derived MWAs and the simulated \UM MWAs so that we can focus solely on the effect of environment on galaxy MWAs, we look at the mean MWA residual from the mean MWA--$M_{\star}$ relation. This allows us to measure the differential effect between inner cluster regions and cluster outskirts (i.e. controlled for stellar mass). In particular, we use interloper galaxies as our reference infalling (already pre-processed) population.

For a simple clean proxy interloper sample that can be easily compared between \UM and SDSS, we choose galaxies in the region of PPS given by
\begin{equation}\label{eq:interloper_selection_in_PPS}
    (1.5<R<2.5, 1.5<V<2) \cup (2<R<2.5, 1<V<1.5)
\end{equation}
as illustrated in Fig.~\ref{fig:interloper_fraction_PPS_distribution} (see also Section~\ref{sec:quenching-model-and-integrated-SFHs}). Based on statistics from the orbit library of \citet{Oman2021}, this region of PPS provides a sample that is made up of $\gtrsim 97$~per~cent true interlopers. Correcting for the $\sim 3$~per~cent impurity is unnecessary for our purposes.

For SDSS, we parameterise the mean MWA--$M_{\star}$ relation (fit and residual shown as black curve in the left panels of Fig.~\ref{fig:MWA_Mstellar}) for (predominantly interloper) galaxies with $R<2.5$ and $V<2$ in the stellar mass range $10^9 < M_{\star}/\mathrm{M}_{\sun} < 10^{12}$ using the function
\begin{equation}
    \overline{\mathrm{MWA}}_\mathrm{SDSS} = A \log \bigg[ \bigg(\frac{M_{\star}}{M_{\mathrm{crit}}}\bigg)^{\alpha} + \bigg(\frac{M_{\star}}{M_{\mathrm{crit}}}\bigg)^{\beta} \bigg] + b, \label{eq:polynomial_MWA_Mstellar_SDSS}
\end{equation}
where $A=0.1$~Gyr is the normalization, $b=9.16$~Gyr is the vertical offset, $M_{\mathrm{crit}}=10^{10.35} \,\Msun$ is the location of the break in slopes, and $\alpha=0$ and $\beta=40$ specify the lower and higher stellar mass slopes, respectively. The fit is chosen to minimize the residual between the running mean MWA--$M_{\star}$ relation and the curve, and an offset is added so that the interloper population has $\mathrm{mean} (\mathrm{MWA}-\overline{\mathrm{MWA}}_\mathrm{SDSS}) = 0$ by construction. This was done rather than fit the interloper MWA-$M_\star$ relation directly since the interloper population had relatively few galaxies to give a good fit across the whole stellar mass range.

For the \UM simulated sample, a polynomial is fit to the $\sim$pure interloper population (Eq.~\ref{eq:interloper_selection_in_PPS}) to the running mean MWA (fit and residual shown as the black curve in the right panels of Fig.~\ref{fig:MWA_Mstellar}). The polynomial fit has the following form:
\begin{align}
    \overline{\mathrm{MWA}}_\mathrm{UM} &= -0.301 + 9.83x - 105 x^2 + 370 x^3; \label{eq:polynomial_MWA_Mstellar_UM}
\end{align}
where $x\equiv \logMstellar$ and the coefficients have units of Gyr.

\subsubsection{Trends in the mean deviation of mass-weighted age in PPS} \label{sec:data-MWA-in-PPS}

\begin{figure*}
	\includegraphics[width=\columnwidth]{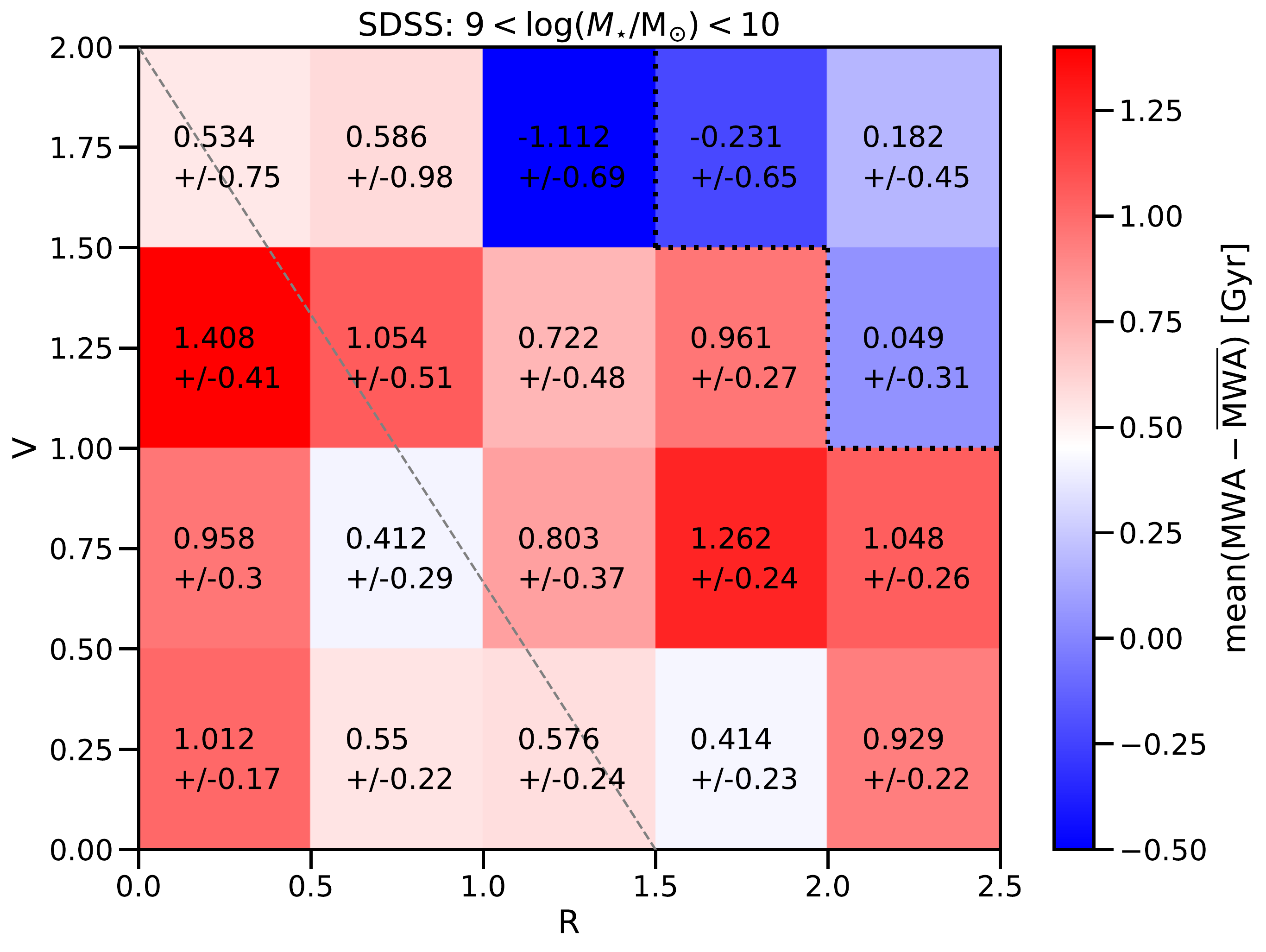}
	\includegraphics[width=\columnwidth]{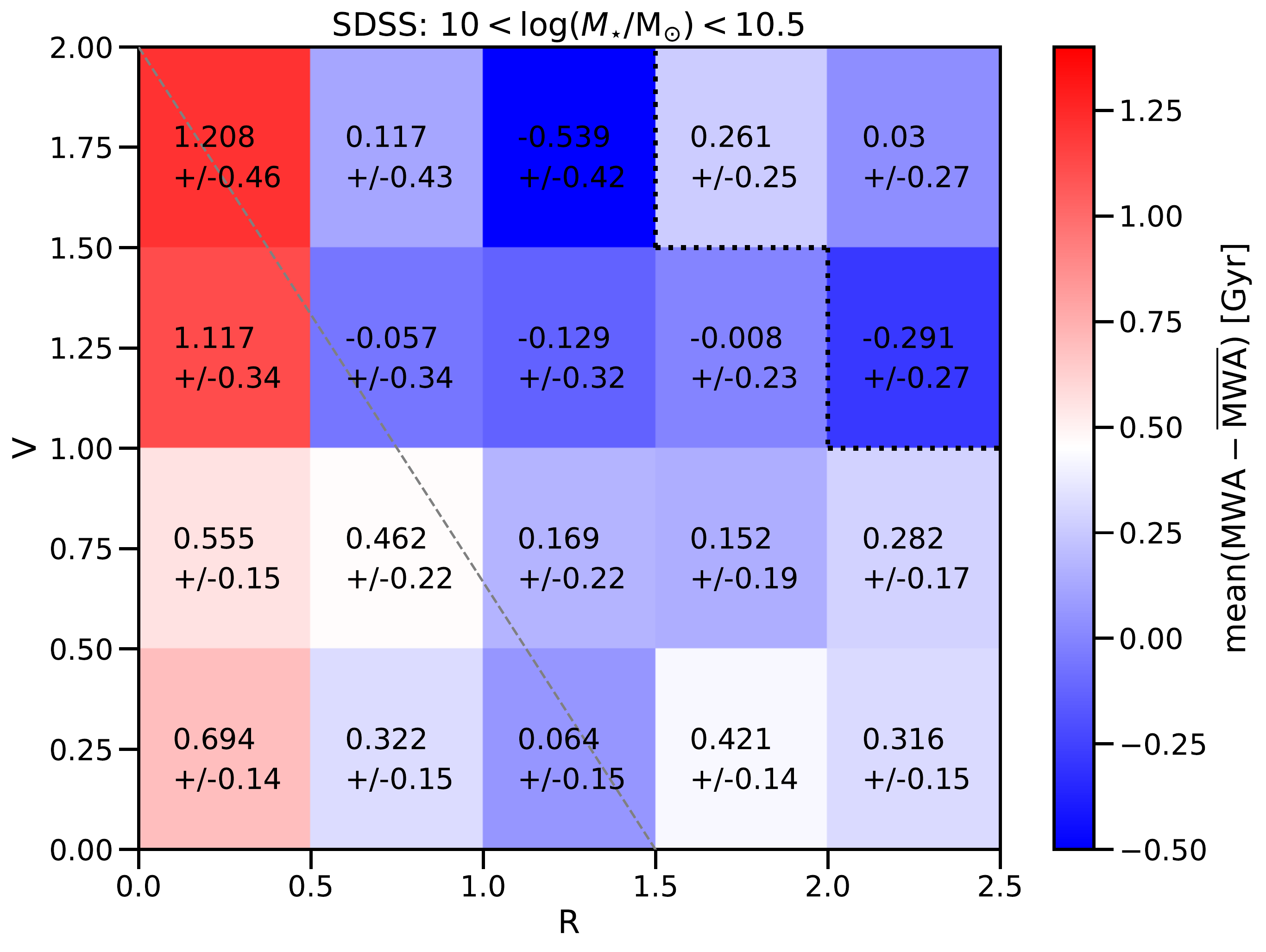}
    \caption{
        Mean deviation in MWA from the MWA-$M_{\star}$ relation of an SDSS interloper proxy sample (top right three bins; outlined by dotted black lines), $\mathrm{mean} (\mathrm{MWA}-\overline{\mathrm{MWA}}_\mathrm{SDSS})$, for quiescent $9<\logMstellar<10$ (left) and $10<\logMstellar<10.5$ (right) SDSS galaxies as a function of PPS coordinates. Errors on the mean values were calculated by bootstrapping over clusters, rather than individual galaxies. We note that the zero-point for this figure, defined by the three PPS bins in the top right corner are 97~per~cent interloper galaxies according the orbit library of \citet{Oman2021}. The dashed grey diagonal line indicates where approximately 50~per~cent of galaxies are interlopers.
    }
    \label{fig:deltaMWA_PPS_distribution}
\end{figure*}

We present a binned map of the deviation in mass-weighted age from the mean MWA--$M_{\star}$ relation of the SDSS interloper proxy sample, $\mathrm{mean} (\mathrm{MWA}-\overline{\mathrm{MWA}}_\mathrm{SDSS})$ (see previous section), as a function of the PPS coordinates $R$ and $V$ in Fig.~\ref{fig:deltaMWA_PPS_distribution}. Errors on the mean values are calculated by bootstrapping over clusters. The figure shows that $\mathrm{mean} (\mathrm{MWA}-\overline{\mathrm{MWA}}_\mathrm{SDSS})$ is enhanced for galaxies at lower radii and velocities. This is particularly clear for the higher stellar mass bin. Within the 50~per~cent interloper line shown in the figure, we calculate $\mathrm{mean} (\mathrm{MWA}-\overline{\mathrm{MWA}}_\mathrm{SDSS})=0.76^{+0.12}_{-0.14}$ and $\mathrm{mean} (\mathrm{MWA}-\overline{\mathrm{MWA}}_\mathrm{SDSS})=0.66^{+0.08}_{-0.09}$ for the lower and higher stellar mass bins, respectively. Our higher stellar mass bin is consistent with \cite{Kim2022}, who find that populations in PPS around clusters at $0.3<z<1.4$ identified as `early infall' were older on average by $0.71\pm 0.4$~Gyr than comparable `recent infall' galaxies.

In the following section we will show that the distribution of modeled $\mathrm{mean} (\mathrm{MWA}-\overline{\mathrm{MWA}}_\mathrm{UM})$, $\Delta$MWA, in PPS differs from that of $f\sbr{Q}$ (shown previously in Fig.~\ref{fig:fQ_SDSS},  Section~\ref{sec:fQ-SDSS}). These differences in observed $f_Q$ and $\Delta$MWA distributions will provide us with a distinct set of preferred values for time of quenching onset and the duration of quenching timescales.

\section{Modelling and Results}\label{sec:modelling-and-results}


Our goal is to model the effects of infall quenching on two SDSS observables, $f\sbr{Q}$ and mean $\Delta$MWA, in PPS around $z\sim 0$ galaxy clusters, and then use these results to constrain the associated quenching timescales. In this section, we first describe our infall quenching model in detail in Section~\ref{sec:our-model}. We show our model results for $f\sbr{Q}$ in Section~\ref{sec:fQ-modelling-results} and for $\Delta$MWA in Section~\ref{sec:ages-modelling-results}. Finally, we combine our modelling results for both $f\sbr{Q}$ and $\Delta$MWA to find a joint constraint on our model's two quenching timescales in Section~\ref{sec:joint-timescale-constraints}.

\subsection{The model} \label{sec:our-model}

To model the effects of infall quenching on both $f\sbr{Q}$ and mean $\Delta$MWA, we use the \UM semi-analytic model, as described in Section~\ref{sec:UM-SFHs}, as a representative model of star formation histories for galaxies in the infall region around a given cluster. A simple, smooth parametric model could be used, but such models don't have any intrinsic scatter in their star formation histories. We will see that this scatter has an impact on the bimodality in the star formation rates of the galaxy population (see Section~\ref{sec:discussion-Wetzel}). The \UM simulation also provides us with a pre-processed infall sample.

Using a galaxy cluster PPS orbit library from \citet{Oman2021}, as described in Section~\ref{sec:oman-orbit-library}, we sample the infall time distribution of galaxies in PPS in galaxy clusters, including the interloper fraction across PPS. With these tools in hand, we describe in Section~\ref{sec:quenching-model-and-integrated-SFHs} how we construct an infall quenching model to predict $f\sbr{Q}$ and $\Delta$MWAs as a function of position in PPS. For the sake of intuition, we visually illustrate some basic trends in $f\sbr{Q}$ and $\Delta$MWA with our quenching parameters in Section~\ref{sec:illustration-of-our-quenching-model}.

\subsubsection{Description of our satellite quenching model} \label{sec:quenching-model-and-integrated-SFHs}

We take the previously defined interloper sample as the infalling population, which will naturally include any pre-processing of galaxies in infalling structures (e.g. groups or filaments), as captured by the \UM model. Our model assumes that star formation histories of $z=0$ cluster satellite galaxies are the same as the $z=0$ infalling population of galaxies, but with the addition of a quenching mechanism on infall, as described below. Galaxies that fall into the cluster begin suppression of their SFR when a time interval of $\tdelay$ has elapsed after their first pericentre passage. Defining the model this way means that we are modeling the mean quenching timescales for the population of galaxies as a whole, over all possible infall orbits (e.g. circular versus eccentric/plunging), giving us robust mean observed properties, insensitive to the choice between stochastic or smooth star formation histories (see Section~\ref{sec:ssfr-bimodality-stochasticity} for discussion of this point). 

For \UM galaxies, we use the same sSFR cut to separate quiescent/star forming galaxies as \citet{Behroozi2019UniverseMachine}, namely 
\begin{equation}
    \log(\mathrm{sSFR/yr}^{-1}) = -11 ,
\end{equation}
where star-forming (quiescent) galaxies are defined as lying above (below) the cut. The \UM interloper population and SDSS interloper population differ slightly in their quiescent fractions (see e.g. Fig.~\ref{fig:adjusted_UM_to_match_SDSS_fQvsMstellar}), but we are only interested in the differential quenching effect ($f\sbr{Q}$ in the cluster versus some infall population). To find the timescales associated with infall quenching, we will fit for the infalling $f\sbr{Q}$. In Appendix~\ref{sec:robustness_adjustSFHs_to_match_SDSS_fQ} we will confirm that our results are robust to adjusting the \UM infalling star formation histories such that the infalling galaxies closely match the SDSS $f\sbr{Q}-M_{\star}$ trend.

In the literature, some variation of a `delayed-then-rapid' quenching scenario is common, like that proposed and explored in \citet{Wetzel2013}. Exponentially declining SFR \citep[e.g.][]{Wetzel2013,Roberts2019} or linearly declining quiescent fraction \citep[such as in][]{Oman2021} are common assumptions. To facilitate comparison with previous work, we choose a model with an exponential suppression of the SFR on a timescale, $\tauenv$, that starts after some time delay, $\tdelay$, relative to the time of first pericentre $\tperi$ (we note that this allows for negative values of $\tdelay$, i.e. quenching onset prior to $\tperi$).

\begin{figure}
	\includegraphics[width=\columnwidth]{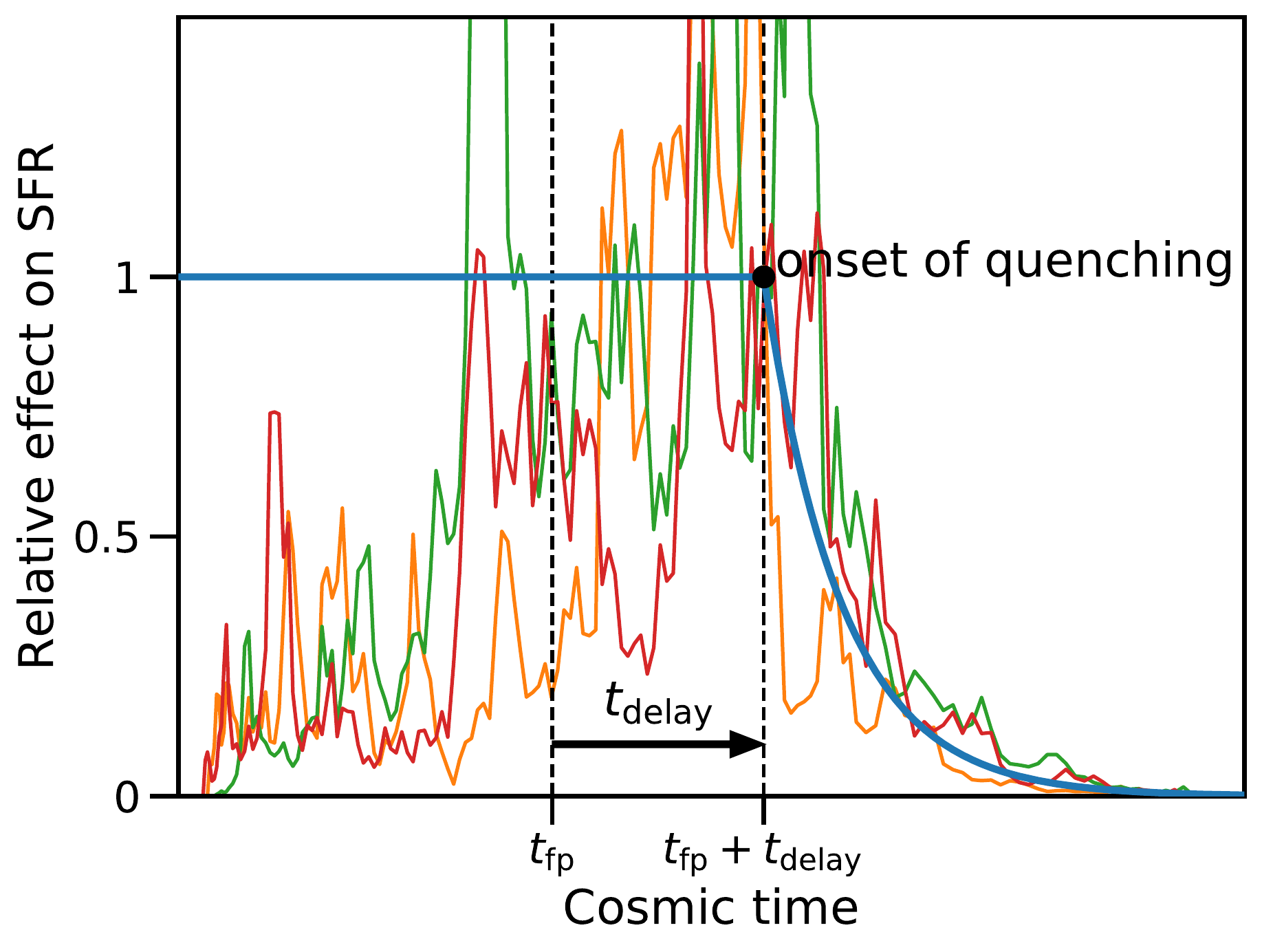}
    \caption{Illustration of the effect of our delayed-then-exponential envelope (thick blue line) on SFR, where at some time delay, $\tdelay$, after time of first pericentre upon infall into a cluster, $t_{\mathrm{fp}}$, SFR is quenched exponentially by a multiplicative envelope with timescale, $\tauenv$.
    }
    \label{fig:quenching_timescales_illustration}
\end{figure}

Galaxies with a given stellar mass and time of first pericentre are Monte Carlo sampled from satellites in the orbital library of \citet{Oman2021}. This gives, for each part of PPS, some mix of interloper galaxies, as well as in-cluster galaxies, which have a distribution of infall times depending on position in PPS. The star formation history for a random interloper with the closest stellar mass to that just drawn is then also drawn from \UM, which gives the star formation history of this Monte Carlo-simulated galaxy. Each of these simulated satellite galaxies have quenching applied by starting the decline in the SFR at the time $\tperi+\tdelay$, i.e. $\mathrm{SFR}_\mathrm{sat}=\mathrm{SFR}_i q(t)$, where $q(t)$ is the multiplicative quenching envelope,
\begin{align}
    q(t; \tdelay, \tauenv) &= 
    \begin{cases}
		1, & t < \tperi+\tdelay, \\
	    e^{-(t-t\sbr{delay}-\tperi)/\tauenv}, & t \geq \tperi+\tdelay,
	\end{cases}
\end{align}
where $i$ refers to a given Monte Carlo-sampled interloper (infalling galaxy) star formation history. In Fig.~\ref{fig:quenching_timescales_illustration}, we illustrate the effect of the multiplicative envelope on the stochastic star formation histories of the \UM galaxies.

We then define our model of the quiescent fraction in a given PPS bin as
\begin{equation}\label{eq:fQ-model}
    f_{\mathrm{Q, model}}(\tdelay, \tauenv, \fQinfall) = \Delta f_{\mathrm{Q}}(\tdelay, \tauenv) + \fQinfall,
\end{equation}
where $\Delta f_Q$ is the increase in quenching from our model and $\fQinfall$ is left as a free parameter of the infalling population's quiescent fraction to account for systematic differences between \UM and the SDSS data.

To determine the stellar masses of any galaxy at $z=0$, the \UM SFRs are integrated numerically according to eq.~21 and eq.~22 in \citet{Behroozi2019UniverseMachine}. Explicitly, this gives a final stellar mass of 
\begin{align}
    M_{\star}(t_{\mathrm{now}}) 
    &= \int_0^{t_{\mathrm{now}}} \frac{\diff M}{\diff t} \diff t \\
    &= \int_0^{t_{\mathrm{now}}} \mathrm{SFR}(t) (1-f_{\mathrm{loss}}(t_{\mathrm{now}} - t)) \diff t ,
\end{align}
where $f_{\mathrm{loss}}(t)=0.05 \ln\big( 1 + t/1.4\mathrm{Myr}\big)$ and $t_\mathrm{now}$ is the age of the Universe at $z=0$.

Integrated true (simulated) stellar masses are offset to give `observed' stellar masses, according to the prescription in \citet[][eq.~25 and eq.~27]{Behroozi2019UniverseMachine}, namely by offsetting them by $\mu = \mathrm{SM}_{\mathrm{obs}} - \mathrm{SM}_{\mathrm{true}} = \mu_0 + \mu_a (1-a)$ [dex] (where $a=1/(1+z)$ is the cosmological scale factor) and additionally adding Gaussian scatter of $\sigma_{\mathrm{SM,obs}} = \mathrm{min}(\sigma_{\mathrm{SM,z}}(1+z), 0.3)$ [dex]. Best-fitting values associated with \UM catalogues available online\footnote{\UM data release 1 catalogues using the \textit{Bolshoi-Planck} dark matter simulation, including catalogues for complete star formation histories, are available on the following page of Peter Behroozi's website: \url{https://www.peterbehroozi.com/data.html}.} were used for these stellar mass adjustments, namely $\mu_0=5.6\times 10^{-3}$ and $\mu_a = -0.03$ used for the offset and $\sigma_{\mathrm{SM,z}}=0.069$ for the Gaussian scatter parameter, respectively.

Mass-weighted ages are integrated numerically from the stellar mass calculations, using the midpoints between the simulation snapshot timesteps and $\diff M / \diff t$ between two snapshots' timesteps,
\begin{equation}
    \mathrm{MWA} = t_\mathrm{now} - \frac{1}{M_{\mathrm{true}}} \int_{0}^{t_{\mathrm{now}}} t \Big( \frac{\diff M}{\diff t} \Big) \diff t ,
\end{equation}
where $M_{\mathrm{true}}$ is the total simulated stellar mass at the present time without any corrections for observational effects. Note that MWA is expressed as a lookback time.

The MWA predicted by our quenching model is then
\begin{equation}
    \mathrm{MWA}_Q = t_\mathrm{now} - \frac{1}{M_{\mathrm{true}}} \int_{0}^{t_{\mathrm{now}}} t \Big( \frac{\diff M}{\diff t} \Big) q(t; \tdelay, \tauenv) \diff t ,
\end{equation}

Similar to $f_Q$ in Eq.~\ref{eq:fQ-model} above, we define the $\Delta$MWA observable of our model in a given PPS bin for each galaxy as
\begin{align}\label{eq:MWA-model}
    \Delta \mathrm{MWA}(\tdelay, \tauenv, \deltaMWA) &= \mathrm{MWA}_{Q,\mathrm{model}}(\tdelay, \tauenv) \\
    &- \overline{\mathrm{MWA}}_\mathrm{UM} \nonumber \\
    &\,+\, \deltaMWA,
\end{align}
where $\mathrm{MWA}_Q(\tdelay, \tauenv)$ is the MWA from our quenching model, $\deltaMWA$ is a floating offset to allow for the possibility that the infalling population's MWA may be different than that of the interloper phase space bins in equation \ref{eq:interloper_selection_in_PPS}, and $\overline{\mathrm{MWA}}_\mathrm{UM}$ is the MWA expected from the mean MWA-$M_\star$ relation as defined in Eq.~\ref{eq:polynomial_MWA_Mstellar_UM} for \UM galaxies. We use the mean $\Delta$MWA in a given PPS bin in the rest of this work.

\subsubsection{Simple illustration of our quenching model}\label{sec:illustration-of-our-quenching-model}

\begin{figure*}
    \includegraphics[width=2\columnwidth]{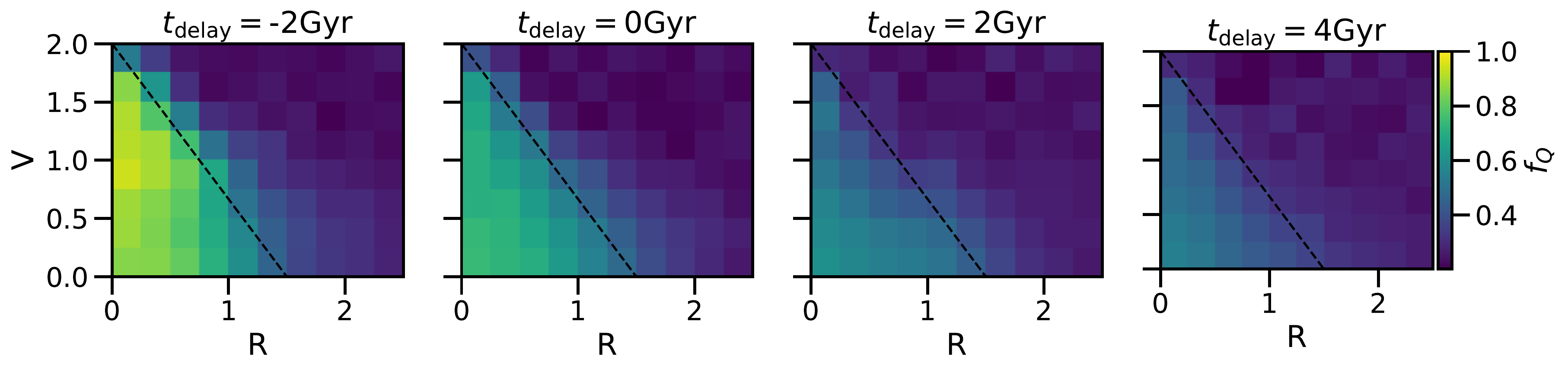}\\
    \includegraphics[width=2\columnwidth]{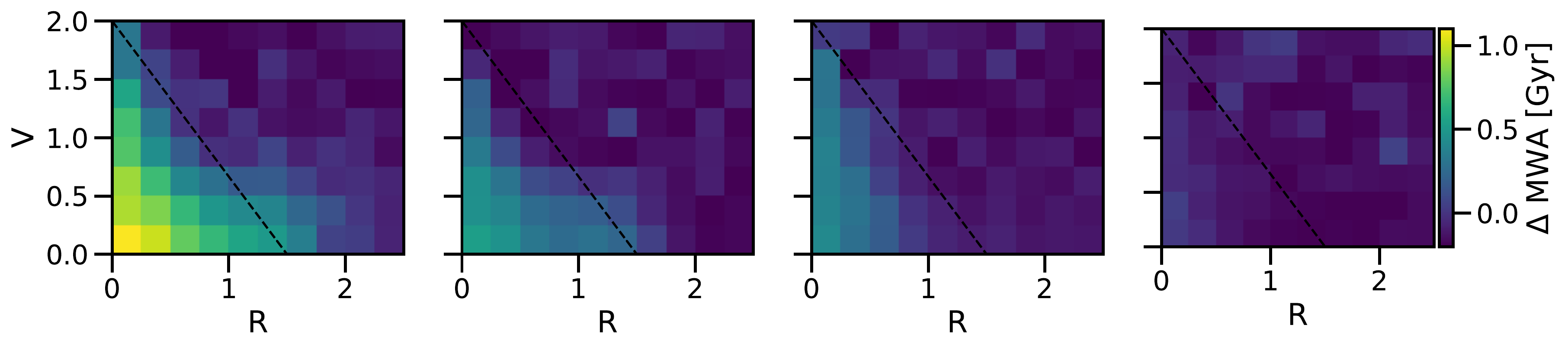}\\
    \caption{
    Delayed-then-rapid (instantaneous quenching, i.e. $\tauenv=0$) model predictions for a range of delay times, $\tdelay$, after time of first pericentre (a negative $\tdelay$ corresponds to quenching prior to $\tperi$). For these illustrative plots, we use the \UM `observed' interloper proxy sample as defined in Section~\ref{sec:MWA_residual_definition}, for clusters with $\logMhalo>14$. \textit{Top row:} predicted quiescent fractions, $f\sbr{Q}$, in PPS. The diagonal dashed line, $V=-(4/3)R+2$, marks the location where $\sim 50$~per~cent of galaxies are interlopers. \textit{Bottom row:} quiescent mass-weighted age predictions in PPS.
    }
    \label{fig:MWA_PPS_tdelay_dependence_tperiModel}
\end{figure*}

Now that we have described our infall quenching model, we illustrate our predictions for trends in $f\sbr{Q}$ and $\Delta$MWA in PPS. In particular, we start by exploring the effect of varying the quenching time delay relative to a satellite galaxy's first pericentre passage, $\tdelay$, assuming instantaneous  quenching (i.e. $\tauenv=0$). We first illustrate this visually in Fig.~\ref{fig:MWA_PPS_tdelay_dependence_tperiModel} for $\tdelay$ ranging from $\tdelay=-2$~Gyr (i.e. prior to first pericentre) to $\tdelay=4$~Gyr. We include contaminating interlopers. 

For $\tdelay=-2$~Gyr, quiescent fractions are very high, $\gtrsim 0.75$ throughout most of the region dominated by satellites (i.e. below the $\sim 50$~per~cent interloper line, indicated by the same diagonal line as in Figs.~\ref{fig:fQ_SDSS} and \ref{fig:deltaMWA_PPS_distribution}). $\Delta$MWA reaches $\sim 1.1$~Gyr for $V<0.5$ in the cluster core ($R<0.5$). $f\sbr{Q}$ and $\Delta$MWA values increase with decreasing radius and also generally increase with decreasing velocity offset (albeit less strongly than with radius).

As $\tdelay$ increases, there is a decrease in both $f\sbr{Q}$ and $\Delta$MWA, as intuitively expected. $f\sbr{Q}$ is still clearly enhanced relative to the interloper population within the $\sim 50$~per~cent interloper line, even for a $\tdelay$ as long as 4~Gyr. $\Delta$MWA reaches $\lesssim 0.25$~Gyr at all $(R,V)$ bins for $\tdelay=4$~Gyr. For an even longer $\tdelay \gtrsim 6$~Gyr (not shown), $f_Q$ continues to drop and $\Delta$MWA drops to $\sim 0$~Gyr.

With this physical intuition in mind, we now carry out a quantitative analysis of our model for $f\sbr{Q}$ and $\Delta$MWA in turn, as well as the joint constraint that observed SDSS $f_Q$ and $\Delta$MWA values place on our model. We will see in the subsections that follow, differences in the trends for these two observables will enable us to break degeneracy in the $\tdelay$ and $\tauenv$ model parameters.

\subsection{Quiescent fraction: comparison of models and data}\label{sec:fQ-modelling-results}

\begin{figure*}
	\includegraphics[width=2\columnwidth]{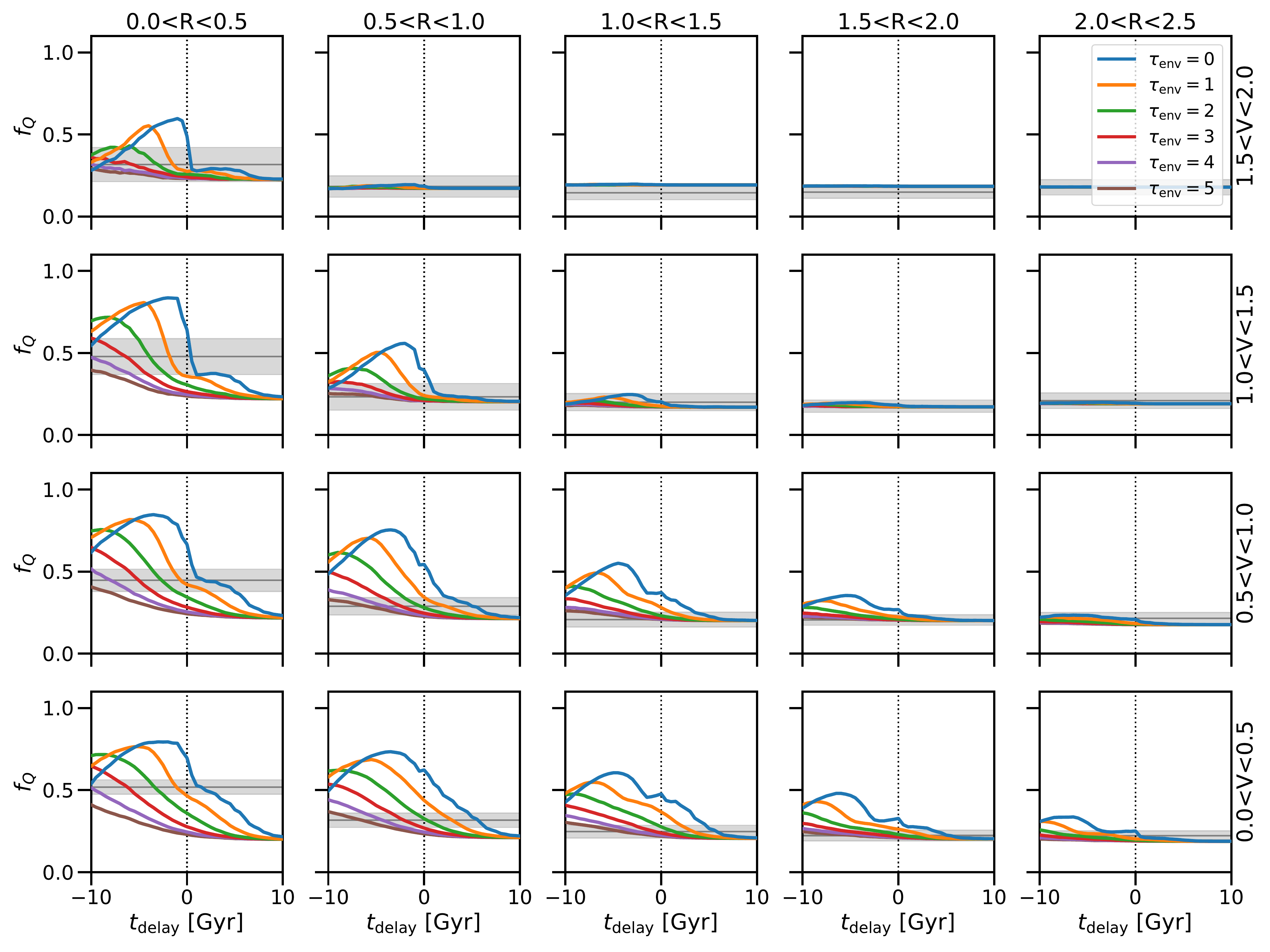}
    \caption{
    Quiescent fraction, $f_\mathrm{Q}$, predictions for $9<\logMstellar<10$ galaxies for a range of models where galaxies quench after some delay time, $\tdelay$, relative to the time of first pericentre. We show models for a range of exponential suppression timescales. The SDSS mean $f_\mathrm{Q}$ values are shown as grey lines, with the shaded regions showing the bootstrapped (over clusters) uncertainty in the mean.
    }
    \label{fig:fQ_tdelay_and_tau_model_lowMstellar}
\end{figure*}

We plot $f\sbr{Q}$ model predictions as a function of time of quenching onset relative to the first pericentre, $\tdelay$, for a range of $\tauenv$ values (different coloured curves) in Fig.~\ref{fig:fQ_tdelay_and_tau_model_lowMstellar} for $9<\logMstellar<10$ galaxies. For the equivalent plot for $10<\logMstellar<10.5$ galaxies at $z\sim 0$, see Fig.~\ref{fig:fQ_tdelay_and_tau_model_midMstellar} in Appendix~\ref{sec:appendix-other-Mstellar-bins}. For reference, we show the mean $f\sbr{Q}$ (grey horizontal line) and the bootstrapped error on the mean (shaded gray region) for the SDSS sample in each PPS bin. When finding the best-fitting $\tdelay$ values (see Section~\ref{sec:joint-timescale-constraints} for this), we also fit for the infalling population's $f\sbr{Q}$ as a nuisance parameter, to account for possible systematic differences between the quenching from \UM and the actual infall population, that we will then marginalize over. For the infalling galaxies, our fits prefer $\fQinfall = 0.16\pm 0.01$ ($\fQinfall=0.48\pm 0.01$) for $9<\logMstellar<10$ ($10<\logMstellar<10.5$) galaxies, shown in Table~\ref{tab:best_fit_parameters}, which are offset $-0.06\pm 0.01$ ($0.09\pm 0.01$) relative to the infall/interloper population's MWAs predicted by the \UM model.

We now turn our attention to the trends in $f\sbr{Q}$ with $\tdelay$ and $\tauenv$, by focussing on the region of PPS where the infall quenching effect in our model is most pronounced, namely the innermost parts of clusters. In e.g. the $R<0.5, V<0.5$ bin we see no effect on $f\sbr{Q}$ for very long quenching times ($\tdelay \sim 10$~Gyr), as expected. $f\sbr{Q}$ then steadily increases with decreasing time delay up until a point at which it begins decreasing again, at e.g. $\tdelay \sim -3$~Gyr. This turnover feature comes about because the star formation histories of infalling galaxies are being truncated so aggressively that many quenched galaxies are dropping out of our stellar mass range, with increasing numbers of quenched galaxies having $\logMstellar<9$. Since there are fewer high stellar mass galaxies, truncated higher stellar mass galaxies dropping into the bin are not enough to compensate for those dropping below $\logMstellar = 9$. The effect is most pronounced for instantaneous quenching, $\tauenv = 0$, and less pronounced for longer star formation suppression timescales, $\tauenv > 0$; $\tauenv$ acts to smooth out dependencies of $f\sbr{Q}$ on $\tdelay$. This delay in quenching due to longer $\tauenv$ values also results in an offset between the maximum possible $f\sbr{Q}$ as a function of $\tdelay$. This offset effect is such that for higher values of $\tauenv$ the maximum possible $f\sbr{Q}$ (the turnover point) requires much earlier quenching, significantly before the time of first pericentre. Features on shorter timescales (i.e. bumps and wiggles) shown in our model curves are due to the discrete snapshot times of the N-body simulation used for the orbit library in \citet{Oman2021}.

We contrast this with the methodology of \citet{Oman2021}, where stellar masses were not truncated by quenching, since for simplicity they did not make use of star formation histories. In that case, there is no turnover effect in $f\sbr{Q}$ as a function of $\tdelay$ since galaxies are not dropping out of the stellar mass range. That said, the $f\sbr{Q}$ data do not prefer excessively negative values of $\tdelay$ and so the shift in preferred delay time compared to our choice of truncating stellar masses is rather minimal: it simply causes an offset of $\sim 0.5$~Gyr (earlier) relative to our model. 

In PPS, the models reproduce the general trends in $f\sbr{Q}$ visible in the SDSS data, but there is clearly significant degeneracy in the preferred model (i.e.\ which $\tdelay$, $\tauenv$ combination is preferred). In order to break this degeneracy, we need another observable beyond the quiescent fraction to constrain $\tdelay$ and $\tauenv$. As mentioned earlier, we will use observationally-derived stellar ages of galaxies for this purpose. We explore model predictions for stellar ages of galaxies in PPS in the next section.

\subsection{Stellar ages: comparison of models and data}\label{sec:ages-modelling-results}

\begin{figure*}
	\includegraphics[width=2\columnwidth]{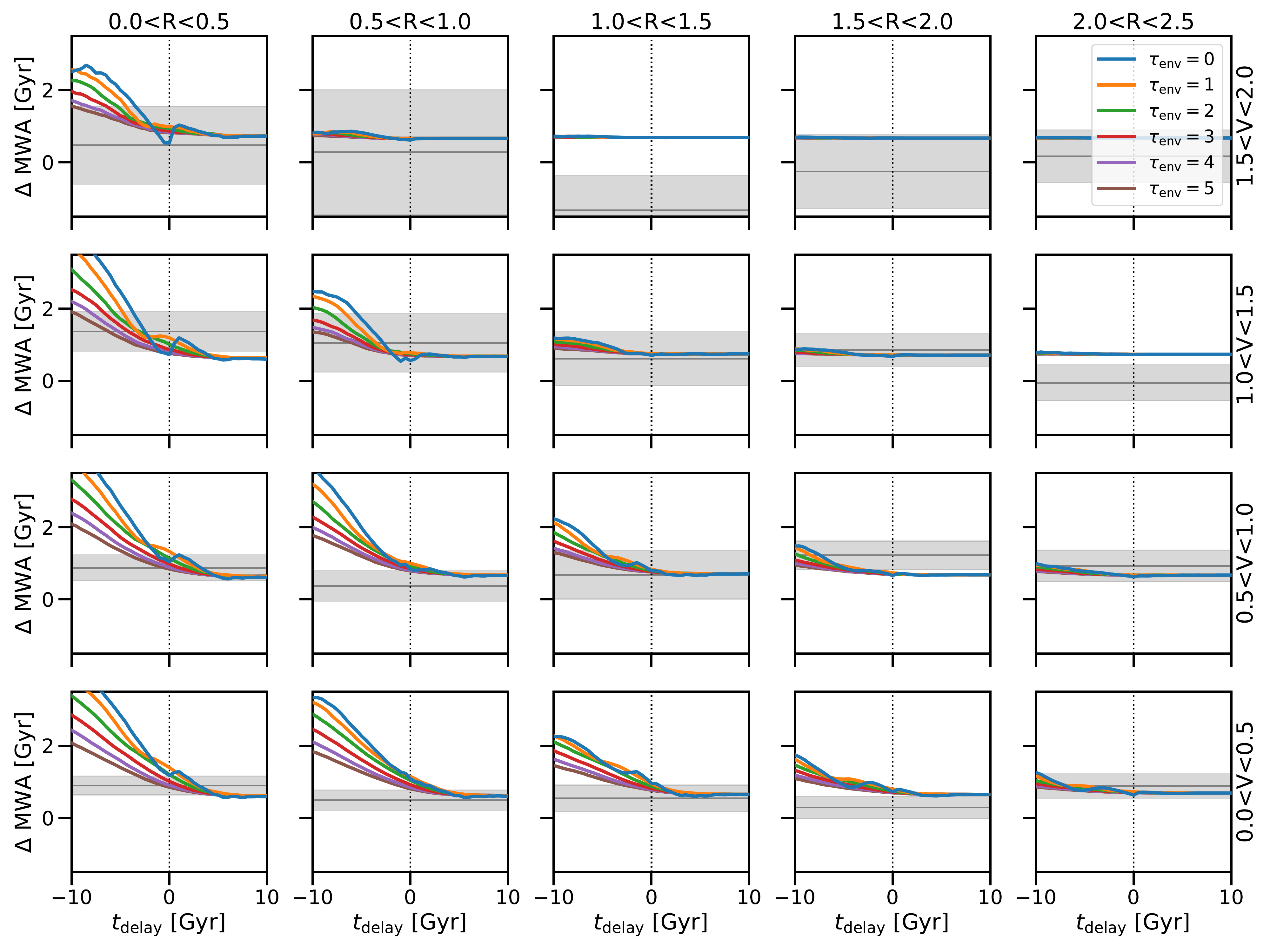}
    \caption{
    Mean $\Delta \mathrm{MWA}$ predictions for quiescent $9<\logMstellar<10$ galaxies for a range of models where galaxies quench after some delay time, relative to time of first pericentre. We show models run for a range of exponential suppression timescales. The SDSS mean $\Delta$MWA values are shown as grey lines, with the shaded regions showing the bootstrapped (over clusters) uncertainty in the mean. Note that our models include an additional offset to allow for the possibility that the infalling population may be different than that of the interloper phase space bins in Eq.~\ref{eq:interloper_selection_in_PPS}.
    }
    \label{fig:deltaMWA_tdelay_and_tau_model_lowMstellar}
\end{figure*}

We now compare our models of stellar age to the SDSS data. As in the previous section, in Fig.~\ref{fig:deltaMWA_tdelay_and_tau_model_lowMstellar}, we plot the mean $\Delta$MWAs as a function of $\tdelay$ in bins of PPS coordinates $R$ and $V$ for $9<\logMstellar<10$. For the equivalent plot for $10<\logMstellar<10.5$ galaxies, see Fig.~\ref{fig:fQ_tdelay_and_tau_model_midMstellar} in Appendix~\ref{sec:appendix-other-Mstellar-bins}. We note that for the infalling galaxies, we find that our fits prefer offsets of $\deltaMWA = 0.59\pm 0.12$~Gyr and $\deltaMWA=0.47\pm 0.07$~Gyr for $9<\logMstellar<10$ and $10<\logMstellar<10.5$ galaxies (shown in Table~\ref{tab:best_fit_parameters}), which correspond to infalling MWAs of $6.70^{+0.11}_{-0.13}$~Gyr and $7.63\pm 0.07$~Gyr.

As in the previous section, looking at e.g. the innermost, low-velocity bin ($R<0.5, V<0.5$), we see no effect on $\Delta$MWA for very long quenching delay times, with the effect on $\Delta$MWA usually increasing with decreasing $\tdelay$ (earlier onset of quenching). This general trend, for the most part, is similar for different values of $\tauenv$, but with shallower slopes for higher values of $\tauenv$. Or, put another way, a longer timescale for the suppression of star formation (a higher value of $\tauenv$) would require an earlier onset of quenching to have the same degree of impact on the stellar age. Interestingly, the trend in higher $\Delta$MWA for earlier onset of quenching relative to time of first pericentre is not monotonic for all values of $\tauenv$: for example, for $\tauenv \sim 0$ (instantaneous or almost-instantaneous suppression of star formation) around $\tperi$ that there is a small $\sim 0.1-0.3$~Gyr dip. This dip is most pronounced for the higher velocity bins, as this is where galaxies reaching their first pericentre will preferentially be located in PPS -- close to the cluster core and moving at a high velocity. Since the figure is only showing the mean $\Delta$MWA of quiescent galaxies, and this region of PPS is dominated by galaxies on their first infall, the influx of recently quenched galaxies lowers the mean $\Delta$MWA relative to values of $\tdelay$ slightly earlier/later than $\tdelay \sim -1$~Gyr (the deepest part of the dip). Longer suppression timescales blur out this effect. Since the lower velocity regions in the cluster core ($R<0.5$) are dominated by galaxies that entered the cluster a long time ago, this effect is washed out.

Although the uncertainties on $\Delta$MWA are larger than for $f\sbr{Q}$, our model still predicts clear trends for $\Delta$MWA across PPS. Because the $\Delta$MWA trends with the timescale parameters differ from those obtained from $f\sbr{Q}$, we explore the constraints that arise from combining $f\sbr{Q}$ and $\Delta$MWA in the next section.

\subsection{Fitting and joint constraints of time delay and exponential quenching timescale}\label{sec:joint-timescale-constraints}

{ 
\renewcommand{\arraystretch}{1.3} 
\begin{table}
\caption{Best-fitting quenching model parameters, given as the 50\textsuperscript{th} percentile for each respective marginal posterior probability distribution. Upper and lower uncertainties are the 84\textsuperscript{th} and 16\textsuperscript{th} percentiles of the marginal distributions. \textit{Top rows:} The time delay until the onset of quenching relative to the time of a galaxy's first pericentre ($\tdelay$), and the exponential star formation suppression timescale after the onset of quenching ($\tauenv$). These are the best-fitting values indicated by the black diamond markers for $\tdelay$ and $\tauenv$ on the subplots of Fig.~\ref{fig:timescales_bestfits_fQplusMWA_break_the_degeneracy}. 
\textit{Bottom rows:} Best-fitting infalling population values, as described for $f\sbr{Q}$ in Eq.~\ref{eq:fQ-model} and $\Delta$MWA in Eq.~\ref{eq:MWA-model}. \textbf{\textdagger} indicates that the value is a 95\textsuperscript{th}-percentile upper limit.}
\label{tab:best_fit_parameters}
\begin{tabular}{ccc}
\hline
Parameter & $9<\logMstellar<10$ & $10<\logMstellar<10.5$ \\
\hline
$\tdelay$ & $3.5^{+0.6}_{-0.9}$ Gyr & $-0.3^{+0.8}_{-1.0}$ Gyr \\
$\tauenv$ & $\leq 1.0$~Gyr$^{\dag}$ & $2.3^{+0.5}_{-0.4}$~Gyr \\ 
\hline
$\fQinfall$ & $0.16\pm 0.01$ & $0.48 \pm 0.01$ \\
$\deltaMWA$ & $0.59\pm 0.12$ & $0.47\pm 0.07$ \\
\hline
\end{tabular}
\end{table}
} 

\begin{figure*}
    \begin{minipage}[t]{0.5\textwidth}
    \centering \large $9<\logMstellar<10$\\
    \includegraphics[width=\textwidth]{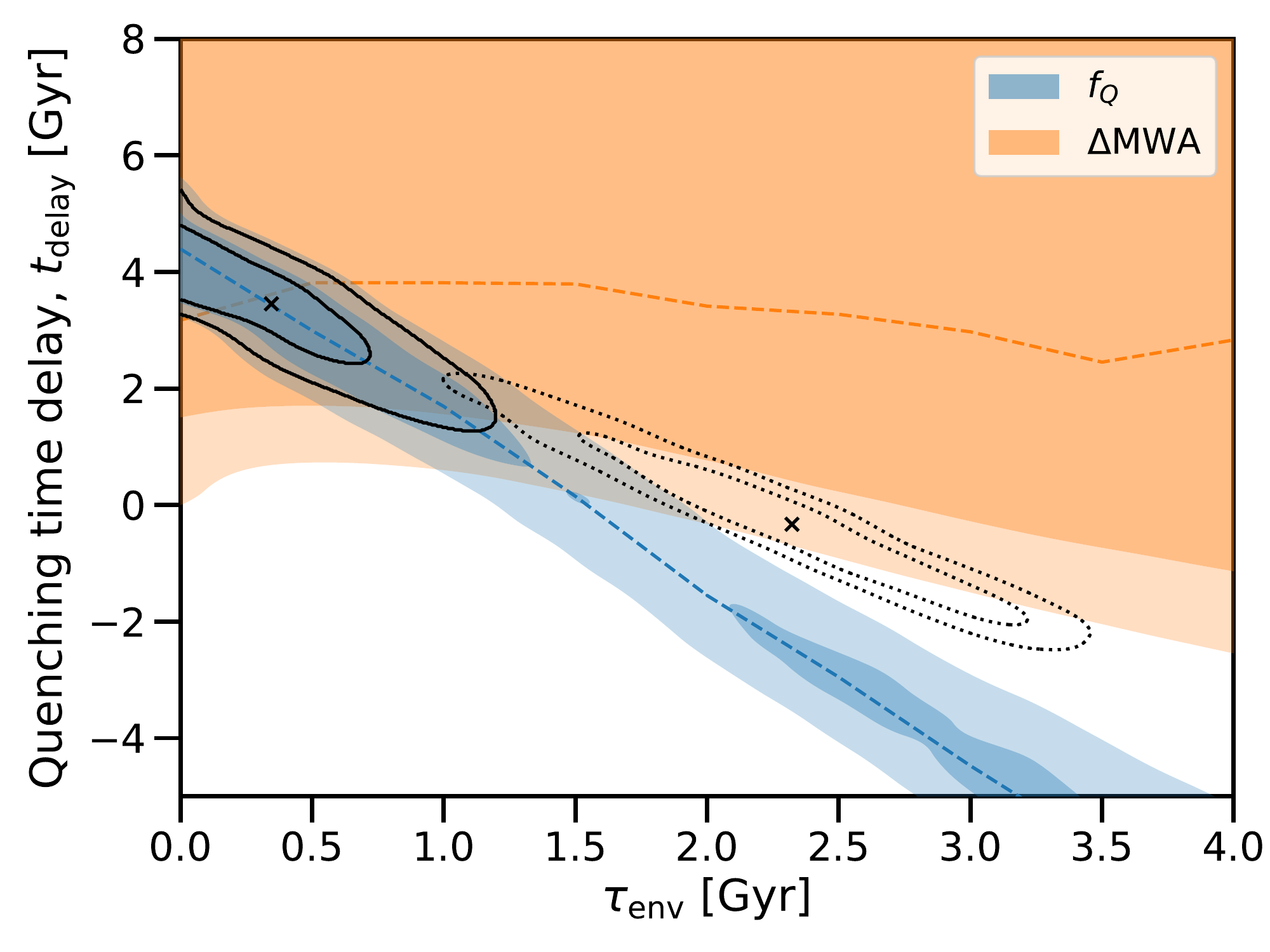}
    \end{minipage}%
    \begin{minipage}[t]{0.5\textwidth}
     \centering \large $10<\logMstellar<10.5$\\
    \includegraphics[width=\textwidth]{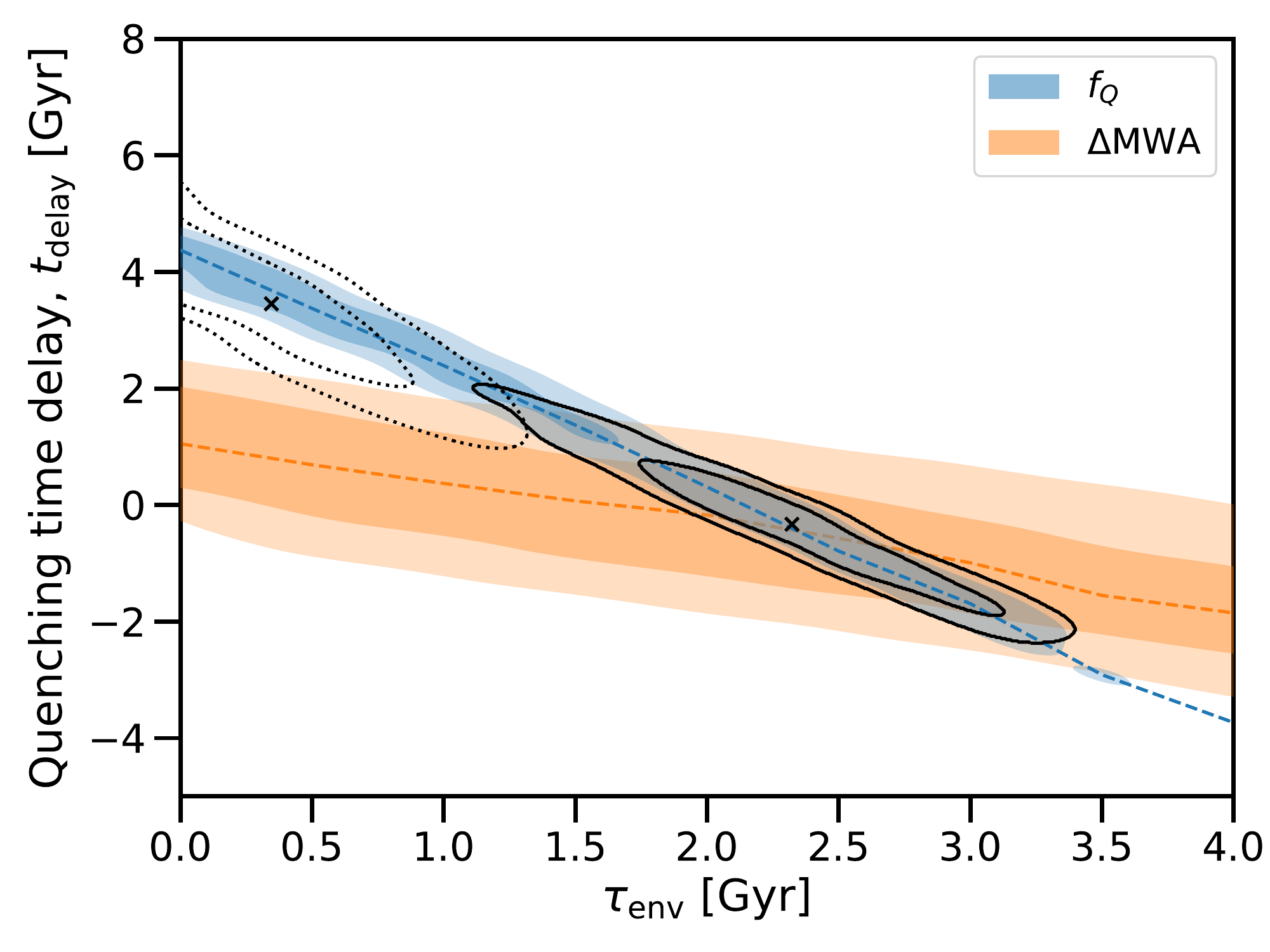}    
    \end{minipage}
    \caption{
    Marginal best-fitting parameter values on our model's quenching timescales for the two observables, $f\sbr{Q}$ (blue), mean $\Delta$MWA for quiescent galaxies (orange), as well as joint (black lines/grey shading) for $9<\logMstellar<10$ galaxies (left) and $10<\logMstellar<10.5$ (right). Uncertainties on the best-fitting parameters are shown with corresponding shaded regions (darker and lighter shaded regions corresponding to the 68\% and 95\% confidence regions, respectively). The parameters fit here are the quenching delay time relative to time of first pericentre, $\tdelay$, and the exponential suppression of star formation timescale, $\tauenv$. Dashed lines indicate the best-fitting $\tdelay$ at a given $\tau$. The joint 68\% and 95\% confidence region from both observables for $\tdelay$ and $\tauenv$ are shown overlaid with black contours. The joint best fitting model, i.e. the location with the peak joint probability, is indicated with a cross symbol. For each plot, the other stellar mass bin's joint constraint is overlaid (black cross and dotted contours). Marginal best-fitting values and uncertainties from the joint constraints are tabulated in Table~\ref{tab:best_fit_parameters}.
    }
    \label{fig:timescales_bestfits_fQplusMWA_break_the_degeneracy}
\end{figure*}

By running our model across a range of $\tdelay$ and $\tauenv$ values we can jointly constrain these timescale parameters by using the information contained in the trends of $f\sbr{Q}$ and $\Delta$MWA as a function of the PPS coordinates. For each $\tauenv$ we perform $\chi^2$ fitting to determine the best-fitting $\tdelay$ value, by minimizing
\begin{equation}
    \chi^2 = \sum_i \bigg(\frac{y_{\textrm{obs},i}-y_{\textrm{model},i}}{\sigma_i}\bigg)^2.
\end{equation}
$y_{\textrm{obs},i}$ and $\sigma_i$ are the mean observed SDSS data value and its uncertainty, respectively, for either $f_Q$ or MWA, in a given PPS bin~$i$. $y_{\textrm{model},i}$ is the modeled value, as defined in Eq.~\ref{eq:fQ-model} and Eq.~\ref{eq:MWA-model}, respectively, and is a function of $\tdelay$, $\tauenv$, and the infall population. The sum is over all PPS bins. We find that the uncertainties on the data may be underestimated, since for the $20-3=17$ degrees of freedom (20 PPS bins, three free parameters) we find the following reduced $\chi^2$ values: $\chi_{\mathrm{red}}^2 (f\sbr{Q}) \approx 2.75$ and $\chi_{\mathrm{red}}^2 (\Delta \mathrm{MWA}) \approx 2.47$ for $9<\logMstellar<10$. Similarly, for $10<\logMstellar<10.5$ we find $\chi_{\mathrm{red}}^2 (f\sbr{Q}) \approx 1.38$ and $\chi_{\mathrm{red}}^2(\Delta \mathrm{MWA})\approx 1.25$. A high $\chi^2$ leads to parameter constraints that may be too tight, so to be conservative we inflate the uncertainties on the observed SDSS mean $f\sbr{Q}$ and mean $\Delta$MWA in each PPS bin by a factor of $\sqrt{\chi_{\mathrm{red}}^2}$. We note that uncertainties on the models are negligible as they can be made arbitrarily small simply by sampling more \UM galaxies.

We present our best-fitting values for $\tdelay$ as a function of $\tauenv$ for our observables $f\sbr{Q}$ and $\Delta$MWA in Fig.~\ref{fig:timescales_bestfits_fQplusMWA_break_the_degeneracy}, for both $9<\logMstellar<10$ (left panel) and $10<\logMstellar<10.5$ (right panel) stellar mass bins. Consistent with the PPS results presented in Section~\ref{sec:ages-modelling-results}, $\Delta$MWA provides a constraint on $\tdelay$ that depends only weakly on $\tauenv$. For $f\sbr{Q}$ there is a steeper trend between the fit $\tdelay$ for a given $\tauenv$. These relations make sense intuitively: as $f\sbr{Q}$ depends on the distributions of SFRs at a given moment in time, it can be greatly impacted by any change in the timescale of SFR suppression, $\tauenv$. $\Delta$MWA, on the other hand, will not change significantly with $\tauenv$ since the bulk of stellar mass growth occurred in the past and will mainly depend on the time at which a galaxy has quenched.

Also shown in Fig.~\ref{fig:timescales_bestfits_fQplusMWA_break_the_degeneracy} are contours showing the confidence intervals for  $\tdelay$ and $\tauenv$ jointly when combining $f\sbr{Q}$ and $\Delta$MWA. We list our best-fitting joint values with uncertainties (16\textsuperscript{th} and 84\textsuperscript{th} percentiles) defined by the $\tdelay$ and $\tauenv$'s marginal probability distributions in Table~\ref{tab:best_fit_parameters}.

Looking first at the $9<\logMstellar<10$ bin in Fig.~\ref{fig:timescales_bestfits_fQplusMWA_break_the_degeneracy} (left panel), quenching time delay relative to time of first pericentre is preferred to be $\tdelay=3.5^{+0.6}_{-0.9}$~Gyr. We find a relatively rapid suppression of star formation once quenching has started, with a best-fitting $\tauenv \leq 1.0$~Gyr (95\% confidence level used as an upper bound). We choose to use a one-sided upper limit for $\tauenv$ for the lower stellar mass bin, as the probability distribution for $\tauenv$ is not Gaussian, peaking at $\tauenv \sim 0$ and the 50\textsuperscript{th}-percentile occurring at $\tauenv = 0.3$~Gyr.

Turning to the higher stellar mass bin, $10<\logMstellar<10.5$ (bottom panel), a different fit is preferred: an earlier onset of quenching, $\tdelay = -0.3^{+0.8}_{-1.0}$~Gyr, corresponding to the onset of quenching beginning close to the time of first pericentre. The corresponding best-fitting $\tauenv$ is $2.3^{+0.5}_{-0.4}$~Gyr, indicating a significantly slower suppression of star formation.


\section{Discussion}\label{sec:discussion}\label{sec:comparisons-to-literature}

We now discuss the meaning of our modelling results and how they contrast with previous literature. For a discussion of the robustness of our results to changes in the overall \UM star formation histories, see Appendix~\ref{sec:robustness_adjustSFHs_to_match_SDSS_fQ}.

We first compare with the work that we have directly built upon, namely that of \citet{Oman2021} \citep[see also:][]{Oman2013orbitLibrary, Oman2016satQuenching}. We then compare our results to the alternative framework proposed in the seminal work of \citet{Wetzel2013}, followed by other observational studies: \citet{Taranu2014} which made use of stellar age-related spectral indices and \citet{Rhee2020} which used PPS information. Finally, we compare to a study examining the predictions of hydrodynamical simulations \citep{Wright2022,Lotz2019}.

\begin{figure*}
    \begin{minipage}[t]{0.5\textwidth}
    \centering \large $9<\logMstellar<10$\\
    \includegraphics[width=\textwidth]{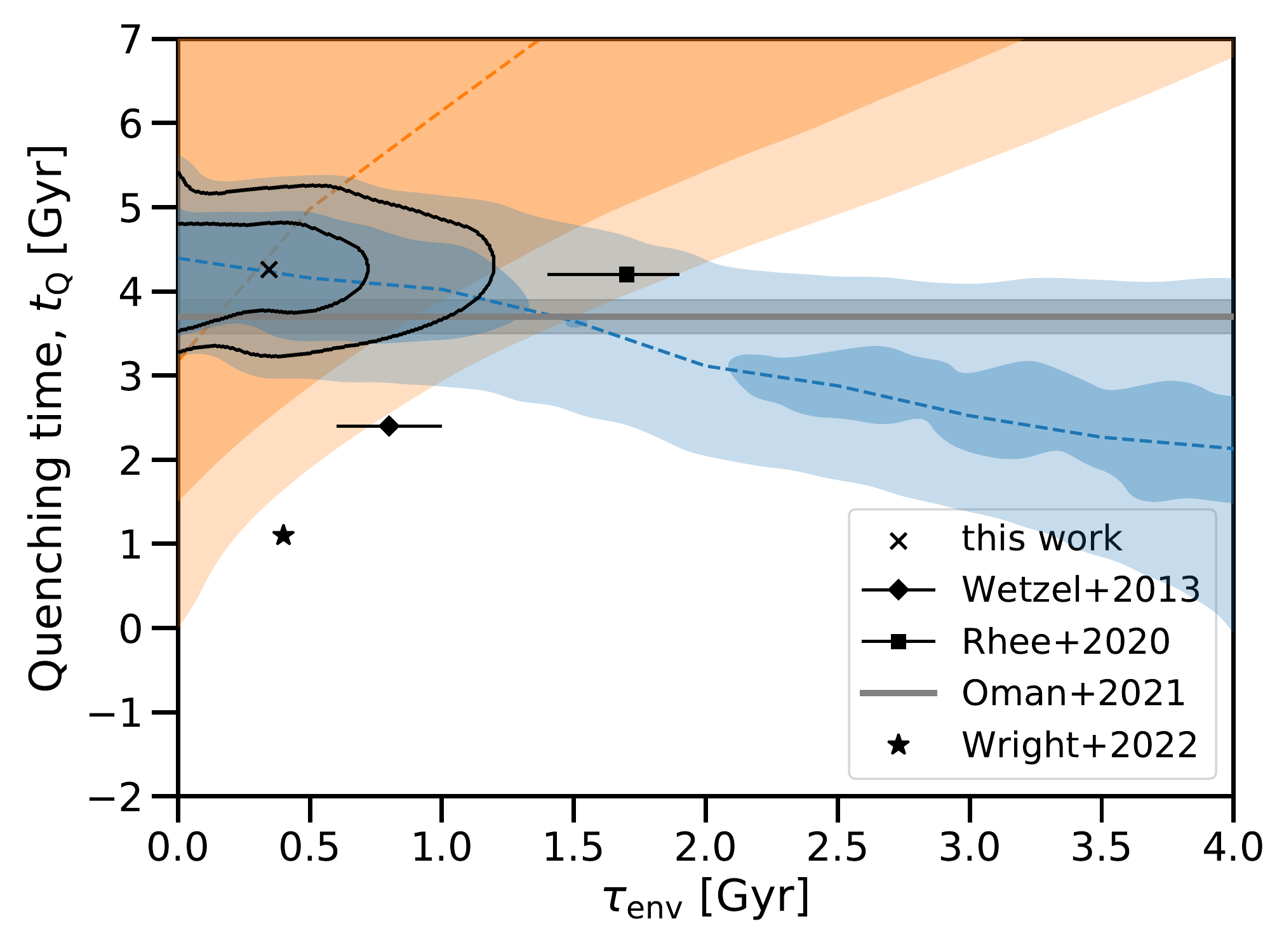}
    \end{minipage}%
    \begin{minipage}[t]{0.5\textwidth}
     \centering \large $10<\logMstellar<10.5$\\
    \includegraphics[width=\textwidth]{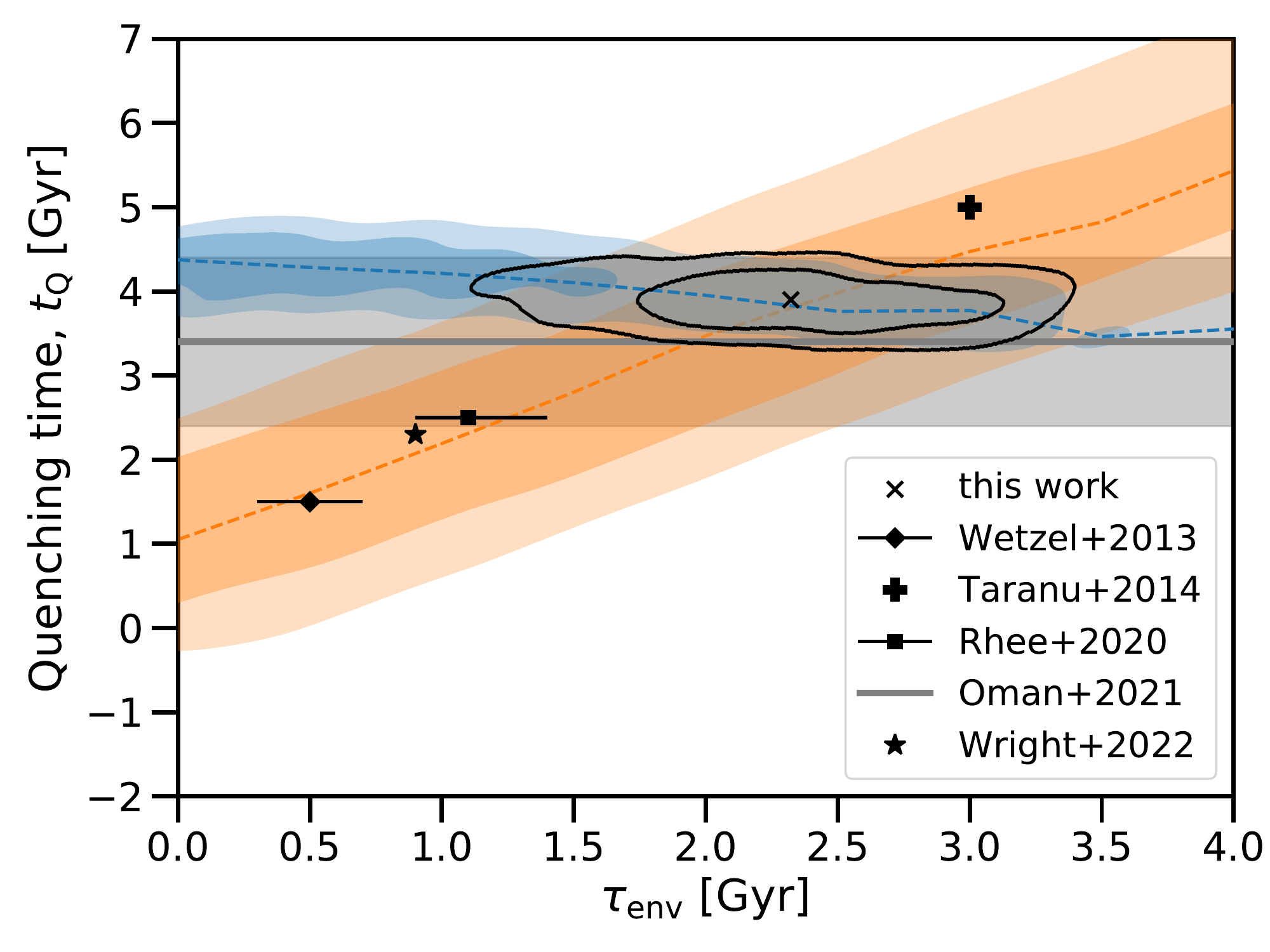}    
    \end{minipage}
    \caption{
    Marginal and joint best-fitting parameter values on overall quenching time, $t\sbr{Q}$, and $\tauenv$ for the two observables, $f\sbr{Q}$ (blue) and quiescent $\Delta$MWA (orange) for $9<\logMstellar<10$ galaxies (left) and $10<\logMstellar<10.5$ (right), analogous to Fig.~\ref{fig:timescales_bestfits_fQplusMWA_break_the_degeneracy}. As well, we have overlaid the $t\sbr{Q}$ literature values from Table~\ref{tab:literature-timescales-comparison} with the various points and lines shown in the legend. We omit error bars where it was not straightforward to determine them.
    }
    \label{fig:tQ_vs_tau_models_plus_literature_values}
\end{figure*}

For the purposes of our discussion, we define a quantity, $t\sbr{Q}$, as the average time for a star-forming galaxy with median sSFR (of a star-forming galaxy) to exponentially decline in SFR until it crosses the \UM quenching threshold of $10^{-11}\,\mathrm{yr}^{-1}$. For our lower (higher) stellar mass bin this difference in sSFR is $\sim 1.0$~dex ($\sim 0.8$~dex). Assuming a constant stellar mass for simplicity, the quenching timescale is $t\sbr{Q} = \tdelay + 2.33\tauenv$ ($t\sbr{Q} = \tdelay + 1.82\tauenv$). These timescales are summarized in Table~\ref{tab:literature-timescales-comparison} and their contours are also plotted in Fig.~\ref{fig:tQ_vs_tau_models_plus_literature_values}. 

\begin{table} 
\centering
\caption{Comparison of timescales from various literature results to our work; in each table section they are listed in the order in which they appear in Section~\ref{sec:discussion}. `Lower' and `higher' stellar mass bins refer to $9<\logMstellar<10$ and $10<\logMstellar<10.5$, respectively. \textit{Upper section:} average time $t\sbr{Q}$ until quenching, relative to time of first pericentre. \textit{ Lower section:} exponential timescales $\tau$ compared to our exponential suppression best-fitting timescale as well as with the conditional assumption ($\bm{*}$) that quenching begins at time of first pericentre ($t\sbr{delay} \equiv 0$), using the $f\sbr{Q}$ contour only. \textbf{\textdagger} indicates use of the upper 95\% confidence level as an upper bound. Note that definitions of $\tau$ in the literature vary; the respective discussion sections should be consulted when making comparisons between $\tau$ values.
}
\begin{tabular}{lc|cc}
\hline
Study & Quantity & Lower $M_{\star}$ bin & Higher $M_{\star}$ bin \\
\hline
\textbf{This work (best)} & \textbf{$\bm{ t\sbr{Q} }$} & $\bm{ 4.3\pm 0.4}$~\textbf{Gyr} & $\bm{ 3.9\pm 0.2 }$~\textbf{Gyr} \\
\textbf{This work ($\bm{t\sbr{delay} \equiv 0})^*$} & \textbf{$\bm{ t\sbr{Q} }$} & $\bm{ 3.7\pm 0.4 }$~\textbf{Gyr} & $\bm{ 4.0\pm 0.2 }$~\textbf{Gyr} \\
Oman+2021 & $t\sbr{Q}$ & $3.7\pm 0.2$~Gyr & $3.4\pm 1.0$~Gyr \\
Wetzel+2013 & $t\sbr{Q}$ & $2.4$~Gyr & $1.5$~Gyr \\
Taranu+2014 & $t\sbr{Q}$ & --- & $5.0$~Gyr \\
Rhee+2020 & $t\sbr{Q}$ & $4.2$~Gyr & $2.5$~Gyr \\
Wright+2022 & $t\sbr{Q}$ & $1.1$~Gyr & $2.3$~Gyr \\
\hline
\textbf{This work (best)} & $\bm{\tauenv}$ & $\bm{ \leq 1.0}$~\textbf{Gyr}$^{\bm \dag}$ & $\bm{ 2.3^{+0.5}_{-0.4}} $~\textbf{Gyr} \\ 
\textbf{This work ($\bm{t\sbr{delay} \equiv 0})^*$} & $\bm{\tauenv}$ & $\bm{ 1.6\pm 0.2 }$~\textbf{Gyr} & $\bm{ 2.2\pm 0.1 }$~\textbf{Gyr}\\
Wetzel+2013 & $\tau_\mathrm{Q,fade}$ & $0.8\pm 0.2$~Gyr & $0.5\pm 0.2$~Gyr \\
Taranu+2014 & $\tau_\mathrm{post}$ & --- & $3$~Gyr \\
Rhee+2020 & $\tau_{\mathrm{cluster}}$ & $1.7^{+0.2}_{-0.3}$~Gyr & $1.1^{+0.3}_{-0.2}$~Gyr \\
Wright+2022 & $\tau$ & $0.4$~Gyr & $0.9$~Gyr \\
\hline
  \end{tabular}
    \label{tab:literature-timescales-comparison}
\end{table}

\subsection{Oman et al. (2021)}\label{sec:discussion-Oman}

The satellite quenching model of \citet{Oman2021}, based on modelling quiescent fractions in PPS, employs a maximum likelihood model constraining four parameters: the quiescent fraction of an infalling population ($f_{\mathrm{before}}$), the final quiescent fraction of galaxies after satellite quenching has taken place ($f_{\mathrm{after}}$, set to zero for their core analysis), the time at which half of the drop in quiescent fraction is complete ($t_{\mathrm{mid}}$, relative to time of first pericentre), and the timescale to go from the initial to final quiescent fraction ($\Delta t$). For the purpose of consistency across the following discussion sections, we will refer to the quenching time as $t\sbr{Q}$ rather than $t_{\mathrm{mid}}$.

We've previously discussed how the preferred time of quenching in \citet{Oman2021} is offset to be $\lesssim 0.5$~Gyr later than ours, since for simplicity they assumed galaxy star formation histories are not truncated by premature quenching.

They find a nearly flat $t\sbr{Q}$ relation with stellar mass (excluding $\logMstellar \gtrsim 10.5$), with $t\sbr{Q}=3.7\pm 0.2$~Gyr and $t\sbr{Q}=3.4\pm 1.0$~Gyr for our lower and higher stellar mass bins, respectively. We find consistent overall quenching times ($t\sbr{Q}(\tdelay=0)=3.7\pm 0.4$~Gyr and $t\sbr{Q}(\tdelay=0)=3.9\pm 0.2$~Gyr, respectively). Relaxing this assumption, our model is still consistent with theirs, although the lower stellar mass bin $t\sbr{Q}$ increases slightly, by $\sim0.5$~Gyr.

\citet{Oman2021} were unable to provide strong constraints on their $\Delta t$ parameter; they find that it tends to be underestimated, e.g. with the median value (based on the probability density from the model) underestimated by up to $\sim 1$ Gyr at lower stellar masses. They note that this effect is due to resolution issues in the N-body simulation resulting in low-mass satellite haloes being disrupted too early. Values of $\Delta t\sim 1-1.5$~Gyr were preferred by their modelling, but with very large uncertainties, which they marginalized over for their primary analysis to get their tight constraints on $t\sbr{Q}$ and the quiescent fraction parameters. They note that the resolution effect could mask a decreasing trend in $t\sbr{Q}$, which could likewise occur in our modelling.

\subsection{Wetzel et al. (2013)}\label{sec:discussion-Wetzel}

\begin{figure}
	\includegraphics[width=\columnwidth]{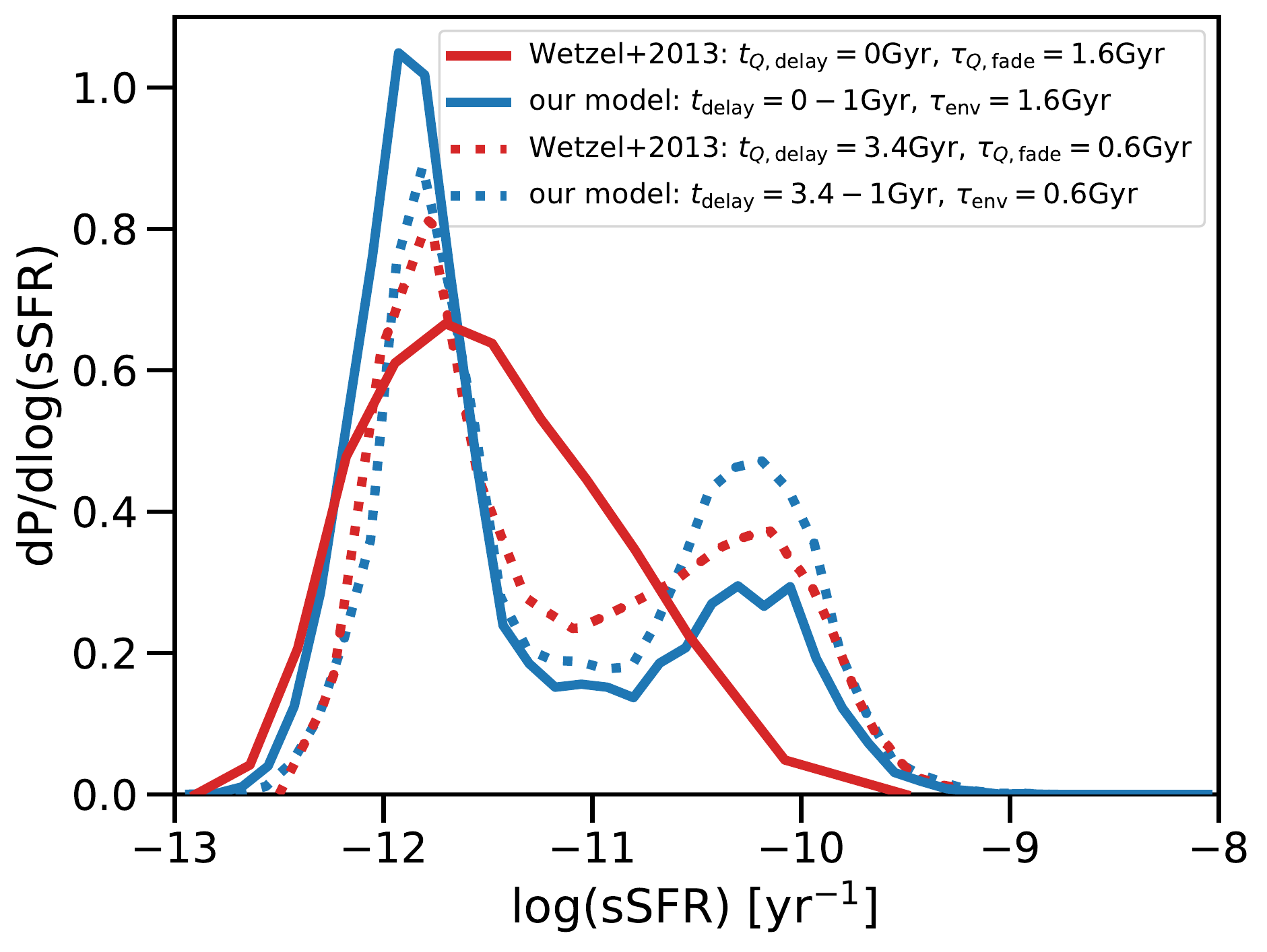}
    \caption{Our model displays a persistent bimodality in sSFR distribution, regardless of SFR suppression timescale, whereas \citet{Wetzel2013} finds a unimodal distribution for long suppression times (contrast the red solid line for their model with the blue solid line for our model). This arises because of our use of a stochastic star formation history, whereas they use smooth star formation histories. In particular, we are contrasting with fig.~9 in \citet{Wetzel2013}, for our $10<\logMstellar<10.5$ stellar mass bin and their $10.1<\logMstellar<10.5$ stellar mass bin. A correction of $-1$~Gyr is added to be comparable to their average infall time, which has the time zero-point set at $r_\mathrm{vir}$. Additionally, for ease of comparison with \citet{Wetzel2013}, who claimed need for a short exponential suppression timescale to reproduce the bimodality, we include only galaxies within $r_{200\mathrm{c}}\approx 0.73 r_{360\mathrm{m}}$ and add log-normal scatter with mean $\log(\mathrm{sSFR}/\mathrm{yr}^{-1})= -12$ and 0.25~dex variance to all galaxies with $\log(\mathrm{sSFR}/\mathrm{yr}^{-1})<-12$.
    }
    \label{fig:Wetzel2013_sSFR_comparison_illustration}
\end{figure}

\begin{figure*}
    \begin{minipage}[t]{0.5\textwidth}
    \centering \large $9<\logMstellar<10$\\
    \includegraphics[width=\columnwidth]{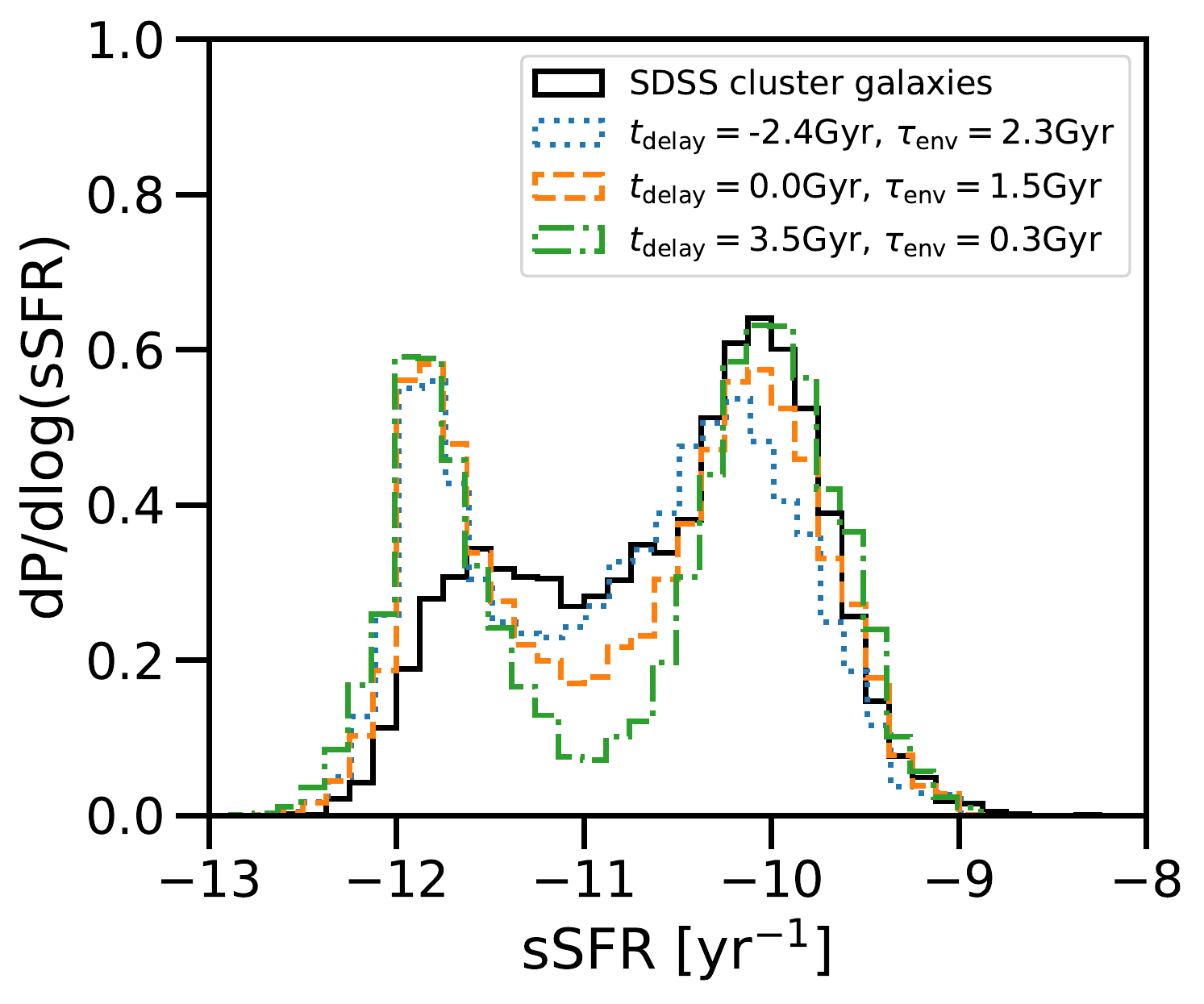}
    \end{minipage}%
    \begin{minipage}[t]{0.5\textwidth}
     \centering \large $10<\logMstellar<10.5$\\
    \includegraphics[width=0.945\columnwidth]{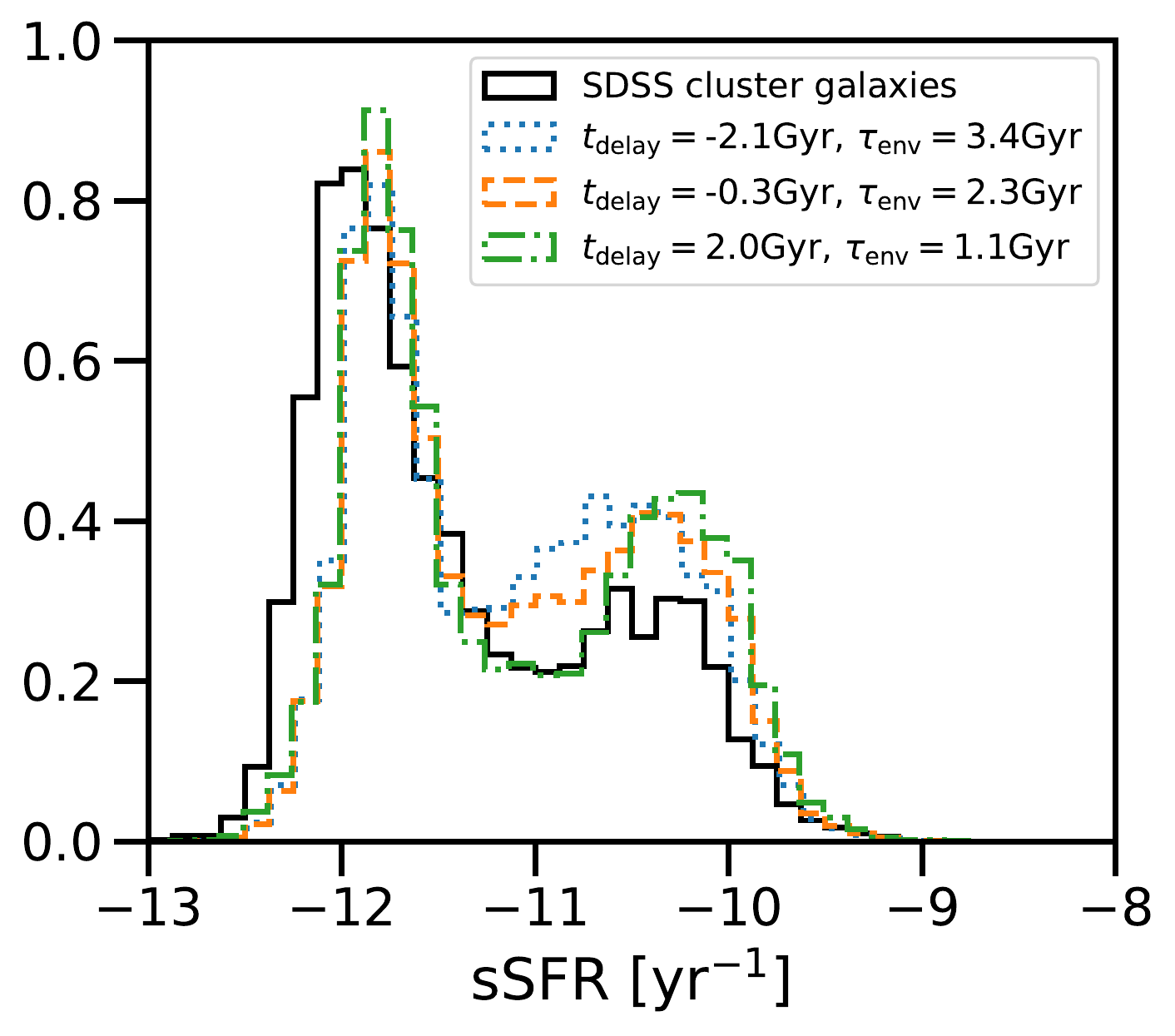}   
    \end{minipage}
    \caption{
    Similar to Fig.~\ref{fig:Wetzel2013_sSFR_comparison_illustration}, except we now demonstrate how we can reproduce the bimodality in the sSFR distributions of the SDSS data (solid black histogram) for a range of relevant models with different exponential quenching timescales (coloured lines) for our two stellar mass bins, $9<\logMstellar<10$ (left) and $10<\logMstellar<10.5$ (right). The left panel shows three scenarios, which are (in the order that they are listed in the legend): (i) long $\tau$ equal to that preferred for our higher stellar mass bin, but $\tdelay$ taken from along the $f\sbr{Q}$ error contour, (ii) a scenario where $\tdelay=0$, with $\tauenv$ chosen to minimize $\chi^2$ (see Fig.~\ref{fig:timescales_bestfits_fQplusMWA_break_the_degeneracy}), (iii) 50\textsuperscript{th} percentile of our timescale parameters. The right panel is similar, showing: (i) rightward end of the 95\% confidence level ellipse, (ii) best-fitting timescale parameters, (iii) leftward end of the 95\% confidence level ellipse. It is worth noting that there is an offset in $f\sbr{Q}$ not factored into the above model histograms, equal to that described in Section~\ref{sec:fQ-modelling-results}, and that the quenching sSFR cut for SDSS has some stellar mass dependence, rather than a simple $\mathrm{sSFR}=10^{-11}\,\mathrm{yr}^{-1}$cut.
    }
    \label{fig:sSFR_distribution_for_various_models}
\end{figure*}

\citet{Wetzel2013} found that satellites falling into a cluster experience `delayed-then-rapid' quenching. One difference between their model and ours is that although \citet{Wetzel2013} fits a delayed-then-rapid model with an exponentially declining sSFR (specifically, their parameters are: time delay, $t_{\mathrm{Q,delay}}$, and `fading timescale', $\tau_{\mathrm{fade}}$), our model relies on the star formation histories of \UM which are stochastic rather than a simple smooth analytical function.

Another important difference in the models is the treatment of pre-processing of galaxies prior to entering the cluster. In our model, a simulated infalling population is used to account for pre-processing, whereas in \citet{Wetzel2013} the infall quenching `clock' for a given galaxy starts on first infall into any larger halo, i.e. when a central becomes a satellite. Because of this, one would expect their (average) quenching time delay to be longer, as their quenching time delay includes the time spent in smaller groups prior to infalling into the main cluster progenitor. The magnitude of the median difference between time of first infall into any halo and infall into the final cluster is $3.2$~Gyr (their fig. 2; assuming $ \log M_{\mathrm{200c}/\mathrm{M}_{\odot}}(z=0)=14.2$ -- our median host halo mass).

\subsubsection{Quenching delay time}

To compare their model with ours, we must account for the two populations of galaxies falling into clusters which will have different quenching times relative to time of first pericentre in the $z\sim 0$ cluster: those that are already satellites in some smaller halo which have had their quenching clock start at some earlier time (A) and those that are falling into a larger halo for the first time (B). For case (A), according to the fits of \citet{Wetzel2013} for a $9<\logMstellar< 10$ galaxy, the full time until quenching for a galaxy that became a satellite in a $\logMhalo>14$ cluster is $\sim 5.7$~Gyr (interpolated). Taking the median time between first infall and infall into the cluster to be $\sim 3.2$~Gyr \citep[see fig.~2 of][]{Wetzel2013} and time from $r_{\mathrm{vir}}$ to first pericentre in the final $\logMhalo>14$ cluster to be $\sim 1$~Gyr \citep[][]{Oman2013orbitLibrary}, we would expect such galaxies to quench $\bar{t}\sim 1.5$~Gyr after first pericentre on average. We note that these pre-processed galaxies are quenching earlier than they would in a scenario where quenching occurs only due to infall into the final cluster host. For case (B), according to their fits for galaxies falling into $\logMhalo>14$ clusters, the full time until quenching is $t\sbr{Q} \sim 4.8$~Gyr for $9<\logMstellar<10$ galaxies, or about $\sim 3.8$~Gyr after first pericentre. The relative proportions of the (A) and (B) scenarios are $\sim 0.6$ and $\sim 0.4$, respectively \citep[see][fig.~1]{Wetzel2013}.

These approximations give an average quenching time, relative to time of first pericentre, of $t\sbr{Q} \sim 0.6(1.5\,\mathrm{Gyr}) + 0.4(3.8\,\mathrm{Gyr})\sim 2.4\,\mathrm{Gyr}$. The same calculation for our higher stellar mass bin gives $t\sbr{Q}\sim 1.5$~Gyr. These are shorter, by $\sim 1.9$~Gyr and $\sim 2.4$~Gyr, than our quenching times of $t\sbr{Q}=4.3\pm 0.4$~Gyr and $t\sbr{Q}=3.9\pm 0.2$~Gyr, for our lower and higher stellar mass bins, respectively.

\subsubsection{Fading time and sSFR bimodality} \label{sec:ssfr-bimodality-stochasticity}

\citet{Wetzel2013} claim that in order for the bimodality in the sSFR distribution to persist across environments, including in the most massive clusters, infall-induced quenching requires a short exponential timescale, $\tau_{\rm{Q,fade}} < 1$ Gyr (they find $\tau_{\rm{Q,fade}}=0.8\pm 0.2$~Gyr and $\tau_{\rm{Q,fade}}=0.5\pm 0.2$~Gyr for our lower and higher stellar mass bins, respectively). For longer SFR fading timescales ($\sim 2$~times their best-fitting $\tau_{\mathrm{fade}}$), their model predicts no bimodality in the sSFR distribution, but rather a unimodal distribution peaking near the centre of the `green valley' ($\log(\mathrm{sSFR})\sim -11.2$ for $\logMstellar<10.5$ galaxies), which we illustrate in Fig.~\ref{fig:Wetzel2013_sSFR_comparison_illustration}. This occurs because their model assumes a smooth exponentially-declining star formation history for the infalling galaxies, and a gentle fading leads to galaxies passing through the green valley too slowly to maintain the bimodality.

We find our lower stellar mass bin exponential timescale ($\tauenv \leq 1.0$~Gyr) is consistent with theirs, but our higher stellar mass bin value of $\tauenv = 2.3^{+0.5}_{-0.4}$~Gyr is significantly longer than their $\tau_\mathrm{Q,fade} \sim 0.5\pm 0.2$~Gyr. However, these fading times are different quantities: $\tau_{\rm{Q,fade}}$ in \citet{Wetzel2013} represents the fading time of an individual galaxy, whereas our $\tauenv$ is a population-wide fading `envelope' which multiplies the stochastic star formation histories of individual \UM galaxies. The stochasticity in star formation histories is due to significant variations in SFR of \UM galaxies, which occur by design: in this model, SFR tracks the accretion rate of baryonic matter, as well having a random variability on short timescales ($\sim 10-100$~Myr). 
This stochasticity allows the sSFR distribution to be bimodal even in the case of a long $\tauenv$. 
Because of the rapid changes in SFR for \UM galaxies, they do not spend significant amounts of time in the green valley, whether or not their SFRs are forced to decline by our exponential suppression envelope (visually illustrated in Fig.~\ref{fig:quenching_timescales_illustration}). The SFRs of infalling galaxies undergoing satellite quenching drop below the sSFR quenched threshold much more abruptly than a suppressed smooth star formation history, so longer fading/suppression envelopes are allowed. 

From a qualitative comparison of the scenarios in Fig.~\ref{fig:sSFR_distribution_for_various_models}, a longer $\tauenv\gtrsim 2.3$~Gyr is preferred by the depth of the green valley for $9<\logMstellar<10$ galaxies, rather than a short $\tauenv \leq 1.0$~Gyr preferred by the $f\sbr{Q}$ and $\Delta$MWA constraints. We note that the location of the star forming peak of the sSFR distribution is better fit by e.g. $\tdelay=0$~Gyr and $\tauenv\sim1.5$~Gyr. For $10<\logMstellar<10.5$ galaxies, we find that the relative depth of the green valley implies a preference for longer $\tauenv$, e.g. $\tauenv \gtrsim 2.3$~Gyr, which is our best-fitting result.

A robust quantitative comparison of infall quenching models using additional sSFR distribution features, aside from $f_Q$, clearly requires controlling for type of star formation history. The depth of the green valley depends on whether the assumed star formation histories are smooth or stochastic and the specific choice of stochasticity made in \UM is not unique. Additionally, measured star formation rates of quiescent galaxies have high systematic uncertainties regardless of how they are obtained \citep{Brinchmann2004,Salim2007,Wetzel2012,Hayward2014}, meaning that much of the shape of the quiescent bump is determined by observational systematics, rather than quenching physics. As such, we do not utilize additional features of the sSFR distribution beyond $f_Q$, as they do not appear informative or robust in constraining our infall quenching model and comparing it to e.g. \citet{Wetzel2013}.

\subsection{Taranu et al. (2014)}

Using an approach broadly similar to our work, \citet{Taranu2014} combined a star formation and quenching model with information from a library of subhalo orbits from N-body simulations of $4$ rich clusters with $\log (M_{200c}/\mathrm{M}_\odot) >15$, with median stellar masses in their sample of $\logMstellar = 10.4$. They compared model predictions to observed bulge and disc colours and are one of the only previous studies \citep[see also][for an attempt with a spectroscopic sample of 11 galaxies]{Upadhyay2021} to make use of age-sensitive stellar absorption line-strength indices for constraining cluster infall quenching times using radial information. Since galaxy disc colours are particularly affected by environment, they can be used to constrain the quenching timescale and location of quenching onset in the cluster using their median trends in cluster-centric radius. They used the age-sensitive Balmer lines of luminous passive cluster galaxies, which are more sensitive to older stellar populations than disc colours.

From their set of models, they found that quenching starting shortly before the time of first pericentre (within a quenching radius of $0.5 r_{200c}$, which is $\sim 0.5$~Gyr prior to the time of first pericentre for a galaxy on first infall) produced the best fits to disc colours as a function of radius. They found that delayed-then-rapid models lead to an excessively large slope in galaxy disc colour versus cluster-centric radius, and also overpredict the strength of the Balmer lines. Instead, their preferred model is one with a gentle exponential quenching timescale of $\tau_\mathrm{post}\sim 3$~Gyr, consistent with the $\tauenv$ that we find for the higher stellar mass bin. This corresponds to a total quenching time of $t\sbr{Q}\sim 5$~Gyr, longer than ours by $\sim 0.7$~Gyr (see Table~\ref{tab:literature-timescales-comparison}).

\subsection{Rhee et al. (2020)}

\citet{Rhee2020} fit a quenching model to disc galaxies in $z=0.08$ clusters. Infall is defined as when a galaxy first crosses $1.5 R_\mathrm{vir}$. They follow an approach similar to ours, parametrizing quenching with a time delay followed by an exponential suppression of star formation, but instead of modelling the quiescent fraction they model the full SFR distribution of galaxies as a function of position in PPS.

They find that the time delay from infall until the onset of quenching is $2$~Gyr for all stellar masses $\logMstellar>9.5$. The time for a galaxy to fall from their infall definition of $1.5 R_\mathrm{vir}$ to first pericentre is $\sim 1.3$~Gyr. We can then subtract this from their quenching time since infall (as presented as $t\sbr{Q}$ in their table~1 for similar stellar mass bins), giving $5.45\,\mathrm{Gyr}-1.3\,\mathrm{Gyr}\sim 4.2\,\mathrm{Gyr}$ and $\sim 3.8\,\mathrm{Gyr (interpolated)}-1.3\,\mathrm{Gyr}=2.5\,\mathrm{Gyr}$ for our lower and higher stellar mass bins, respectively. This is consistent with our lower stellar mass bin (we find $t\sbr{Q}\sim 4.3 \pm 0.4$~Gyr) but earlier for our higher stellar mass bin by $\sim 1.4$~Gyr (our $t\sbr{Q} \sim 3.9 \pm 0.2$~Gyr).

This difference in trend with stellar mass between our results and those of \citet{Rhee2020} is not due to a difference in delay times, but rather is due to their SFR suppression timescale, $\tau_\mathrm{cluster}(z=0)$, which declines with increasing stellar mass over our stellar mass range, whereas we find the opposite trend. We note that they add a redshift-dependent factor to their quenching timescale $\tau_\mathrm{cluster}(z_\mathrm{inf})=\tau_\mathrm{cluster,0}(1+z)^{-\alpha}$ and to the quenching time delay timescale (due to the redshift evolution of the dynamical time in clusters), $t_\mathrm{d}(z_\mathrm{inf})=t_{\mathrm{d},0} (1+z_\mathrm{inf})^{-1.5}$. This only results in a small correction (a decrease by $\sim 10$~per~cent if we use e.g. $z=0.5$, the time at which an average galaxy has fallen into the cluster). For $z=0$, they find $\tau_{\mathrm{cluster,0}}=1.7^{+0.2}_{-0.3}$~Gyr and $\tau_{\mathrm{cluster,0}}\sim 1.1^{+0.3}_{-0.2}$~Gyr for our lower and higher stellar mass bins, respectively. This is consistent with our lower stellar mass bin's $\tauenv(\tdelay=0) = 1.6\pm 0.2$~Gyr, but more rapid than our higher stellar mass bin's $\tauenv(\tdelay=0)=2.2\pm 0.1$~Gyr, respectively.

\subsection{Comparison with hydrodynamical simulations} \label{sec:discussion-hydro-sims}

A study directly comparable to ours, which examines quenching in the EAGLE hydrodynamical simulations, is that of \citet{Wright2022}. Their work provides an orbital analysis of galaxies' gas inflow, stripping, star formation and quenching. Stripping of infalling galaxies' hot gas begins at 2-3 virial radii from the host and takes longer for high-mass satellites ($\logMstellar>10$). This begins the process of starvation and the removal of the hot gas `buffer', resulting in galaxies becoming vulnerable to cold gas stripping. They refer to the hot gas halo as having a protective effect, as they observe that the onset of significant cold gas stripping only begins after stripping of the hot gas halo is complete. In their work, they include both \HI and molecular hydrogen in the mass of cold gas in a galaxy. 

For low-mass satellites ($\logMstellar<10$), they find that suppression of gas cooling onto the galaxy becomes permanent after hot gas stripping, normally occurring around the time of first pericentre. Some high-mass satellites, on the other hand, retain small hot gas reservoirs, and continue to cool gas for star-formation after first pericentre. Cold gas stripping is shown to be periodic, being strongest for galaxies near pericentre, as that is when density and velocity is at a maximum, hence maximizing the ram-pressure force ($P_{\mathrm{ram}}\propto \rho_{\textsc{icm}} v^2$). All of this results in the following: low mass satellites experience very efficient ram-pressure stripping of cold gas, leading to rapid quenching, whereas high mass satellites experience less efficient stripping and a more gradual starvation-like scenario after their first pericentre.

In terms of quenching, we note that \citet{Wright2022} remove pre-processed infalling satellites, namely those galaxies which were satellites in a host halo with $\log (M_{200}/\mathrm{M}_{\odot}) \geq 12$ prior to falling into the current host ($\approx 30$~per~cent of infalling galaxies). This handling of pre-processing is somewhat different from ours, as we fit the pre-processed quiescent fraction for infalling galaxies and focus on the differential quenching (relative to the infalling population) in the final $z=0$ cluster.

To compare our quenching timescales with \citet{Wright2022}, we compare the time required for a galaxy of average sSFR (see their fig.~7) to cross the quenching threshold and assume $\tdelay=0$ to find an approximate average quenching time relative to the time of first pericentre. \citet{Wright2022} finds median quenching times $0.25 \langle T_{\mathrm{orb}}\rangle = 0.25 (4.5\,\mathrm{Gyr})=1.1$~Gyr and $0.65 \langle T_{\mathrm{orb}} \rangle = 0.65 (3.5\,\mathrm{Gyr})=2.3$~Gyr relative to the time of first pericentre, for their $9<\logMstellar<10$ and $10<\logMstellar<11$ bins, respectively. For the higher stellar mass galaxies, the majority ($\sim 80$~per~cent) are quenched by second pericentre. We find longer quenching times of $t\sbr{Q}=3.7 \pm 0.4$~Gyr for $9<\logMstellar<10$ galaxies and $t\sbr{Q}=4.0 \pm 0.2$~Gyr for $10<\logMstellar<10.5$ galaxies. Using similar reasoning, their SFR suppression is equivalent to an exponential suppression time of $\tau \sim 0.4$~Gyr and $\sim 0.9$~Gyr for our lower and higher stellar mass bins, respectively. Taking $\tdelay=0$, our $\tauenv$ is significantly longer, by $1.2$~Gyr and $1.3$~Gyr, respectively. Relaxing $\tdelay=0$, we find little to no change in preferred $t\sbr{Q}$, but results in our preferred exponential suppression timescale being consistent with theirs for lower stellar mass galaxies.

Based on this discussion, we conclude that we prefer significantly longer total quenching timescales than \citet{Wright2022}. If we assume $\tdelay=0$, as their models predict, then we find that our star formation suppression timescales, $\tauenv$, are longer than theirs by $1.2-1.3$Gyr.

A similar previous study is that of \citet{Lotz2019}, which instead examined $z\sim 0$ quiescent fractions in the Magneticum Pathfinder hydrodynamical simulation. They found that most $\logMstellar<10.5$ galaxies are quenched within $\sim 1$~Gyr of crossing $r_{200c}$ (i.e. around the time of first pericentre), with the relatively small fraction of galaxies with tangential orbits and very high stellar masses able to maintain star formation after first pericentre. This quenching is significantly earlier than we find for both of our stellar mass bins, and all of the results examined in our discussion above.

\subsection{Towards a consistent model of quenching}

There is ample observational evidence that ram pressure stripping of the cold gas starts at or just before first pericentre. For example, \cite{SmithLuceyHammer2010} found that ram-pressure stripped tails of Coma cluster galaxies were prevalent within half of the cluster virial radius and that most of these tails pointed away from the cluster, indicating that the stripping was occurring on infall, i.e. just before pericentre. Studies of the distribution of \HI abundance as a function of the PPS coordinates \citep{Jaffe2015, Oman2021} find that \HI depletion begins close to pericentre, i.e within $0.5 r_{200c}$, in agreement with \cite{SmithLuceyHammer2010}. However, quenching is not instantaneous and likely proceeds from outside inwards if it is due to ram-pressure stripping \citep{BoselliFossatiSun2022}. \cite{Owers2019} studied cluster galaxies with spatially resolved spectroscopy and uncovered a population of galaxies with strong H$\delta$ absorption -- indicative of recent quenching -- in their outskirts, but these same galaxies had ongoing star formation in their centres. They modeled the distribution of these galaxies in PPS, finding indications that galaxies with a recent quenching event in their outskirts are within 1 Gyr of entering within $0.5 r_{200c}$ of the cluster centre.

That quenching should start at (or just before) pericentre is supported by our results for our higher stellar mass bin ($\tdelay = -0.3\pm^{+0.8}_{-1.0}$~Gyr) but is at odds with our results for low stellar mass galaxies. For these we find that the onset of quenching occurs well past pericentre $\tdelay = 3.5\pm^{+0.6}_{-0.9}$~Gyr and with a fast quenching envelope ($\tauenv \leq 1.0$~Gyr). Such a short quenching timescale, however, would predict a deep `green valley' in the sSFR distribution, inconsistent with the observed shallow depth (see Fig.\ \ref{fig:sSFR_distribution_for_various_models}). For the lower stellar mass bin, quenching starting at pericentre ($\tdelay=0$) is permitted by the quiescent fraction but disfavoured by the stellar ages at the $\gtrsim 2 \sigma$ level (see Fig.~\ref{fig:timescales_bestfits_fQplusMWA_break_the_degeneracy}). This preference is driven by there being little-to-no gradient in $\Delta$MWA between galaxies in the cluster core and those infalling, as shown in Fig.~\ref{fig:deltaMWA_PPS_distribution}. This result, however, appears to be driven partly by the very lowest stellar mass galaxies: if we restrict analysis to $9.5<\logMstellar<10$, we find somewhat better compatibility with $\tdelay = 0$. If we fix $\tdelay=0$, as suggested by other observational and theoretical evidence, then, for the lower stellar mass bin, we find slower quenching with $\tau\sbr{env} = 1.6\pm0.2$ Gyr. This timescale is in better agreement with the sSFR distribution in Fig.\ \ref{fig:sSFR_distribution_for_various_models}. For the higher stellar mass sample, the quenching timescale is longer, with $\tau\sbr{env} = 2.2\pm 0.1$~Gyr, and is in reasonable agreement with the sSFR distribution. While the SFR suppression timescales that we find are longer than those found in hydrodynamical simulations (see Section~\ref{sec:discussion-hydro-sims}), they are shorter than the gas depletion timescales of 3.5--4 Gyr for field galaxies of comparable stellar mass \citep{BoselliCorteseBoquien2014}.

Taking all of these results together suggests a picture in which ram pressure stripping starts close to pericentre and is effective in a satellite galaxy's outskirts (where the restoring force is low compared to the force due to ram pressure) but may not be fully effective in their more tightly bound inner regions. For gas in the inner regions, while there may be no inflow of new cold gas (due to the complete stripping of the hot gas in the halo), the remaining cold gas will then be consumed by star formation. This consumption timescale is shorter than in the field for two reasons: first, because there is less cold gas available due to the stripping in the outskirts; and second, because ram pressure stripped galaxies have modestly-enhanced star formation rates at fixed stellar mass, compared to similar galaxies in the field, of 0.2--0.3 dex \citep{RobertsParker2020, RobertsParkerGwyn2022}.


\section{Conclusions}\label{sec:conclusions}

We have combined SDSS photometry and spectroscopy, orbital information from tracking haloes in an N-body simulation, and simulated galaxies from the \UM empirical model to constrain a simple model that suppresses star-formation histories of galaxies that have fallen into clusters. In the model, an infalling galaxy will have its star formation rate suppressed by an exponential envelope with timescale $\tauenv$ after some delay time $\tdelay$, relative to the time of the first pericentre. The parameter fits from this modeling give the mean quenching timescales for the infalling population as a whole. We jointly fit these model parameters using both the quiescent fraction and mean deviation from the mean MWA-$M_{\star}$ relation ($\Delta$MWA) of the infalling population in projected phase space. Doing so allows us to break the degeneracy between time of quenching onset and quenching duration that was present in previous models. The method accounts for interloper galaxies (which appear in projection, but are not physically in the cluster), and the pre-processing of infalling galaxies, allowing us to isolate the quenching effect of the (most recent) infall into a massive cluster. Our main results can be summarized as follows:
\begin{itemize}
    \item The mean mass-weighted stellar age depends on location in projected phase space, with cluster member-dominated regions being older (by $\lesssim 1$~Gyr) relative to an interloper-dominated region.
    \item Overall quenching times for our two stellar mass bins are driven by the quiescent fraction and are consistent with \citet{Oman2021}, whose methodology we build on directly: $t\sbr{Q} = 4.3\pm 0.4$~Gyr and $t\sbr{Q} = 3.9 \pm 0.2$~Gyr for our lower and higher stellar mass bins, respectively. We find longer overall quenching timescales than other works in the literature where only star-formation rates are modeled, but agree with \cite{Taranu2014}, who make additional use of the age-sensitive Balmer lines of quiescent galaxies.
    \item Using mass-weighted ages allows us to break the degeneracy between $\tdelay$ and $\tauenv$. We find that the onset of quenching occurs at $\tdelay = 3.5^{+0.6}_{-0.9}$~Gyr and $\tdelay = -0.3^{+0.8}_{-1.0}$~Gyr, relative to time of first pericentre, for galaxies in our $9<\logMstellar<10$ and $10<\logMstellar<10.5$ stellar mass bins, respectively. The models prefer a short SFR suppression timescale, $\tauenv \leq 1.0$~Gyr (consistent with ram-pressure stripping), for our lower stellar mass bin, and a longer $2.3^{+0.5}_{-0.4}$~Gyr (consistent with strangulation) for our higher stellar mass bin.
    \item In contrast to \citet{Wetzel2013}, our model is able to reproduce the SFR bimodality even with long exponential suppression timescales, thanks to the stochasticity of the \UM star formation histories that we employ (as opposed to using smooth analytic star formation histories). We note that, for our lower stellar mass bin, the depth of the green valley prefers values of $\tauenv \gtrsim 1.5$ Gyr and $\tdelay\sim 0$, in slight tension with the later quenching onset preferred by $\Delta$MWAs.
\end{itemize}

Based on these findings and on our detailed discussion of the literature, we argue that satellites infalling into clusters experience ram pressure stripping of cold gas starting close to pericentre, which is only effective in the galaxy's outskirts, at least on the first pericentre passage. This leaves reduced cold gas available for continued star formation, resulting in star formation suppression timescales of $\tauenv\sim 2$~Gyr -- longer than if galaxies were fully stripped on their first pericentre passage, but shorter than a simple starvation scenario.

Future surveys like the Bright Galaxy Survey of the Dark Energy Spectroscopic Instrument (DESI) will provide a large increase in sample size over that of the one million galaxies in the SDSS DR14 main galaxy sample \citep{RuizMacias2021}. An increase in sample size of galaxies with spectroscopically-derived MWAs should reduce the errors on $\Delta$MWA significantly. Using spectroscopically-derived quantities more sensitive to recent SFR suppression or quenching, such as the time at which 90~per~cent of the stellar mass has formed \citep{Webb2020,Upadhyay2021}, could also provide additional constraining power on infall-related and general quenching models. 
The models could also be improved by using a physically-motivated model of ram-pressure stripping, rather than a generic timescale for SFR decline. Such a model could involve radius and velocity at first pericentre, as suggested in a simple model by \citet[][see also \citealp{Roberts2019} for a quenching model depending on ICM density]{Owers2019}. 
With these various improvements, tighter constraints on time of quenching onset and duration via infall quenching as a function of stellar mass should be possible in the future.

\section*{Acknowledgements}

We thank the anonymous referee for their helpful comments. Addressing these clarified a number of key points for understanding this work. We thank S.~Ellison for assistance with SDSS stellar masses and sSFRs. AMMR acknowledges support from a Queen Elizabeth II Graduate Scholarship in Science and Technology (QEII-GSST) from the province of Ontario. MJH acknowledges support from a NSERC Discovery Grant. KAO acknowledges support by the European Research Council (ERC) through Advanced Investigator grant to C.~S.~Frenk, DMIDAS (GA 786910), and by STFC through grant ST/T000244/1. This research has made use of NASA's Astrophysics Data System.

This work used the DiRAC@Durham facility managed by the Institute for Computational Cosmology on behalf of the STFC DiRAC HPC Facility (www.dirac.ac.uk). The equipment was funded by BEIS capital funding via STFC capital grants ST/K00042X/1, ST/P002293/1, ST/R002371/1 and ST/S002502/1, Durham University and STFC operations grant ST/R000832/1. DiRAC is part of the National e-Infrastructure.

The University of Waterloo acknowledges that much of our work takes place on the traditional territory of the Neutral, Anishinaabeg and Haudenosaunee peoples. Our main campus is situated on the Haldimand Tract, the land granted to the Six Nations that includes six miles on each side of the Grand River. Our active work toward reconciliation takes place across our campuses through research, learning, teaching, and community building, and is centralized within the Office of Indigenous Relations.

Funding for the Sloan Digital Sky Survey (SDSS) has been provided by the Alfred P. Sloan Foundation, the Participating Institutions, the National Aeronautics and Space Administration, the National Science Foundation, the U.S. Department of Energy, the Japanese Monbukagakusho, and the Max Planck Society. The SDSS Web site is \url{www.sdss.org}.

The SDSS is managed by the Astrophysical Research Consortium (ARC) for the Participating Institutions. The Participating Institutions are The University of Chicago, Fermilab, the Institute for Advanced Study, the Japan Participation Group, The Johns Hopkins University, Los Alamos National Laboratory, the Max-Planck-Institute for Astronomy (MPIA), the Max-Planck-Institute for Astrophysics (MPA), New Mexico State University, University of Pittsburgh, Princeton University, the United States Naval Observatory, and the University of Washington.

\section*{Data Availability}

The VVV simulation's initial conditions and snapshots are not currently publicly available. If they are needed, please ask the VVV authors for them. Alternatively, any N-body simulation of reasonably similar resolution and cosmology should yield the same results, statistically speaking. For satellite orbit and interloper data tables based on this simulation, request them from KAO (\href{mailto:kyle.a.oman@durham.ac.uk}{kyle.a.oman@durham.ac.uk}). Alternatively, they may be created for any N-body simulation using the publicly available \textsc{rockstar} (\href{http://bitbucket.org/gfcstanford/rockstar}{bitbucket.org/gfcstanford/rockstar}), \textsc{Consistent Trees} (\href{http://bitbucket.org/pbehroozi/consistent-trees}{bitbucket.org/pbehroozi/consistent-trees}) and ORBITPDF (\href{http://github.com/kyleaoman/orbitpdf}{github.com/kyleaoman/orbitpdf}) codes.

The \UM $z\sim 0$ simulation results using the \textit{Bolshoi-Planck} dark matter simulation, including full star formation histories for all galaxies, are available in catalogue form via download from \href{https://www.peterbehroozi.com/data.html}{peterbehroozi.com/data.html}.

The SDSS data used in this work is derived from \citet{Oman2021} and is available publicly as follows: Data Release 7 used in this work is available at \href{http://skyserver.sdss.org/dr7}{skyserver.sdss.org/dr7}; catalogues with stellar masses are available from VizieR (\href{http://vizier.u-strasbg.fr}{vizier.u-strasbg.fr}), catalogue entry \texttt{J/MNRAS/379/867}; SFRs are available from \href{https://wwwmpa.mpa-garching.mpg.de/SDSS/}{wwwmpa.mpa-garching.mpg.de/SDSS/}; the two group catalogues are separately available through VizieR (catalogue entry \texttt{J/MNRAS/379/867}) and \href{http://gax.sjtu.edu.cn/ data/Group.html}{gax.sjtu.edu.cn/data/Group.html}. The mass- and luminosity-weighted ages and their accompanying stellar mass estimates are available from \href{https://www.sdss.org/dr16/spectro/galaxy\_firefly}{sdss.org/dr16/spectro/galaxy\_firefly}.



\bibliographystyle{mnras}
\bibliography{bibliography} 



\appendix

\section{Robustness of parameter constraints to changes in star formation history and floating infall parameters} \label{sec:robustness_adjustSFHs_to_match_SDSS_fQ}

An essential assumption for our modelling results and conclusions to be robust is that they are not particularly sensitive to changes in overall star formation history, for both $f\sbr{Q}$ as well as $\Delta$MWA. Throughout this work, we simply allowed our infalling/interloper \UM galaxies' $f_Q$ to float to match that of SDSS. As a test of these assumptions, we now consider a modified model where the \UM star formation histories are adjusted such that the infalling \UM galaxies have the same quiescent fraction as for SDSS.

The mismatch between the level of quenching observed in the infall region of \UM and SDSS clusters is particularly noticeable at stellar masses $\logMstellar > 10$. To be specific, we now check whether we need to have star formation histories that match the $z\sim 0$, $f\sbr{Q}-M_{\star}$ relation of infalling \UM galaxies to that found for SDSS. For the purposes of this test we consider interlopers to be galaxies in the three upper-right bins in our PPS plots, namely the bins that define the infall region in e.g. Fig.~\ref{fig:interloper_fraction_PPS_distribution}. 

\begin{figure*}
	\includegraphics[width=2\columnwidth]{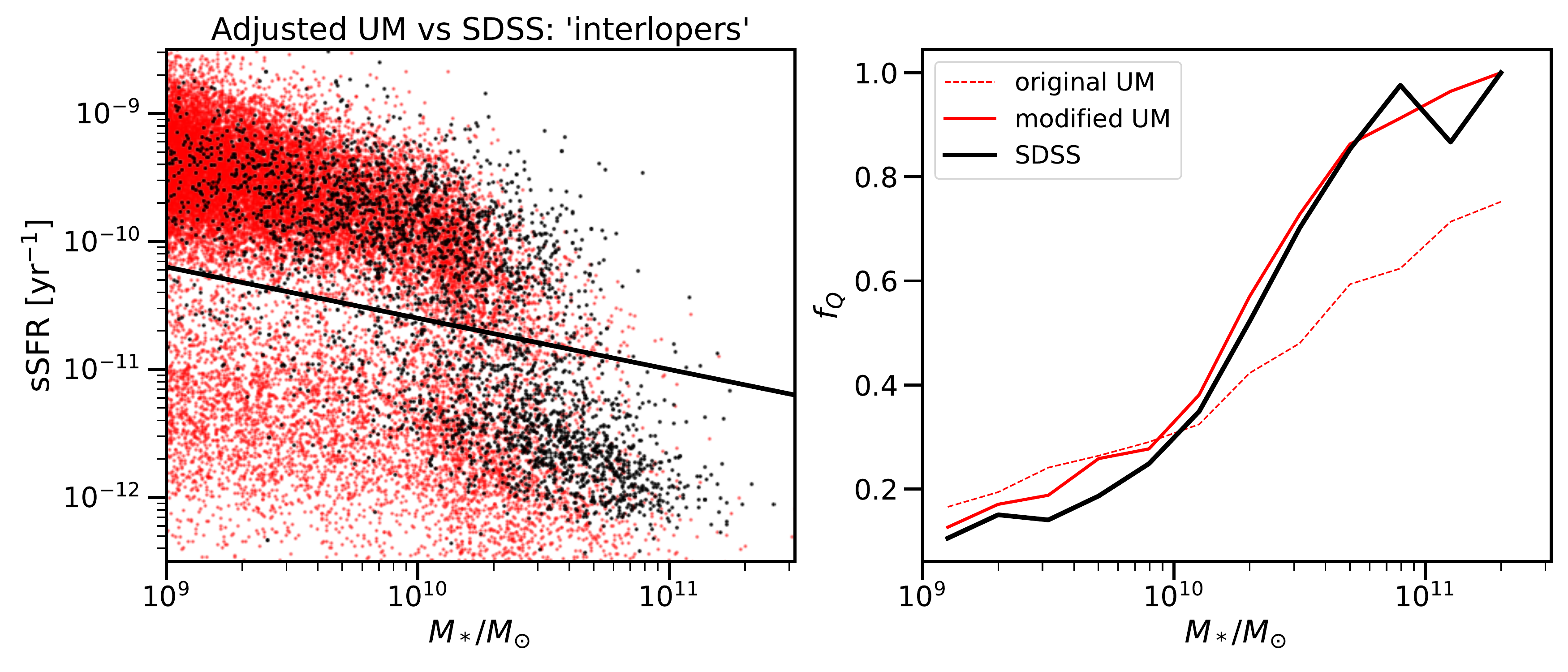}
    \caption{Plots illustrating the adjusted \UM interlopers (red) relative to the SDSS data (black). Above $\logMstellar=10.2$, galaxies have had their star formation history suppressed to increase their $f\sbr{Q}$ to match the SDSS data, whereas galaxies below $\logMstellar=10.2$ have had their star formation boosted. \textit{Left:} specific star formation rate plotted against stellar mass, with the adjusted \UM sSFR multiplied by 2.1 (vertical shift) for the purposes of comparing with SDSS data on this plot. The quiescent/star-forming cut used for both is shown with the solid black line. \textit{Right:} quiescent fraction versus stellar mass. The adjusted \UM interloper sSFRs increase their $f\sbr{Q}$ (dashed red) up to that of the SDSS data for higher stellar masses. There is a small offset in the two curves as only the shape of the $f\sbr{Q}-\mathrm{M}_*$ relation was fit in this stellar mass binning and the absolute $f\sbr{Q}$ values were fit for stellar mass bins with edges at $\logMstellar=(9,10,10.5,11,11.5)$.
    }
    \label{fig:adjusted_UM_to_match_SDSS_fQvsMstellar}
\end{figure*}

Our quiescent fraction is several per cent too high at low stellar masses and several per cent too low at higher stellar masses, as apparent in Fig.~\ref{fig:adjusted_UM_to_match_SDSS_fQvsMstellar}. Our floating $\fQinfall$ parameter removes this overall offset, leaving behind just the differential trends (across PPS) in $f\sbr{Q}$. This resolves possible issues due to differences in completeness between star forming and quiescent galaxies, particularly in the lower stellar mass bin. However, as a robustness check to see how much differences in $f\sbr{Q}$ might affect our measured timescales, we run a test where we modify the \UM star formation histories such that we closely match the $f\sbr{Q}-M_{\star}$ trend observed in SDSS. Star formation histories are suppressed by multiplying by an exponential function in time such that the \UM sSFR-$M_{\star}$ slope better matches the SDSS data and also better matches the SDSS $f\sbr{Q}$. Explicitly, this modified star formation history in terms of the unmodified \UM star formation history (SFH) as a function of cosmic time, $t$, is expressed as:
\begin{equation}
    \mathrm{SFH}'(t) = \exp{\left(\frac{t-t_{z=0}}{\tau}\right)} \, \mathrm{SFH}(t),
\end{equation}
where $t_{\mathrm{now}}=13.8$~Gyr, $\tau = 1/k$, and $k=(\ln(10)/t_{\mathrm{now}}) \times -0.4(\logMstellar - 10)$, with $-0.4$ coming from the slope of the SDSS sSFR cut. The value of this correction at $z=0$ is such that the increase or decrease in SFR forms a broken power law with slope $-1$ for $\logMstellar>10$ and slope $-0.7$ for $\logMstellar<10$, with the value of the correction set to 1.0 at $\logMstellar=10.3$. This corresponds to a boost in the star formation histories for $\logMstellar<10.3$ and suppression for $\logMstellar>10.3$. The correction is applied to the `true' rather than `observed' \UM SFHs as well as the `true' final SFRs and stellar masses \citep[see][for these definitions]{Behroozi2019UniverseMachine}. A shift plus scatter is then applied to approximate observed versions of the quantities afterwards, according to the prescription used in \citet{Behroozi2019UniverseMachine}.

We find that this adjustment only results in minor effects on the best-fitting $\tdelay$ and $\tauenv$ parameters. Specifically, the preferred $\tdelay$ changes by $+0.2$~Gyr and $\tauenv$ by $-0.2$~Gyr for our lower stellar mass bin, or in terms of $t\sbr{Q}$, a shift of $0.5$~Gyr. For the $10<\logMstellar<10.5$ stellar mass bin, the corresponding shifts in $\tdelay$ and $\tauenv$ are $+0.1$~Gyr and $-0.5$~Gyr, respectively. 
Given our uncertainties of $\sim 0.5$~Gyr for $\tauenv$ and $\lesssim 1.0$~Gyr for $\tdelay$ 
(see Table~\ref{tab:best_fit_parameters}), we conclude that any impact on our parameter constraints from changes in overall star formation history is negligible and therefore does not influence our discussion and qualitative conclusions.


\section{Detailed modelling predictions for higher stellar mass bin} \label{sec:appendix-other-Mstellar-bins}

\begin{figure*}
	\includegraphics[width=1.75\columnwidth]{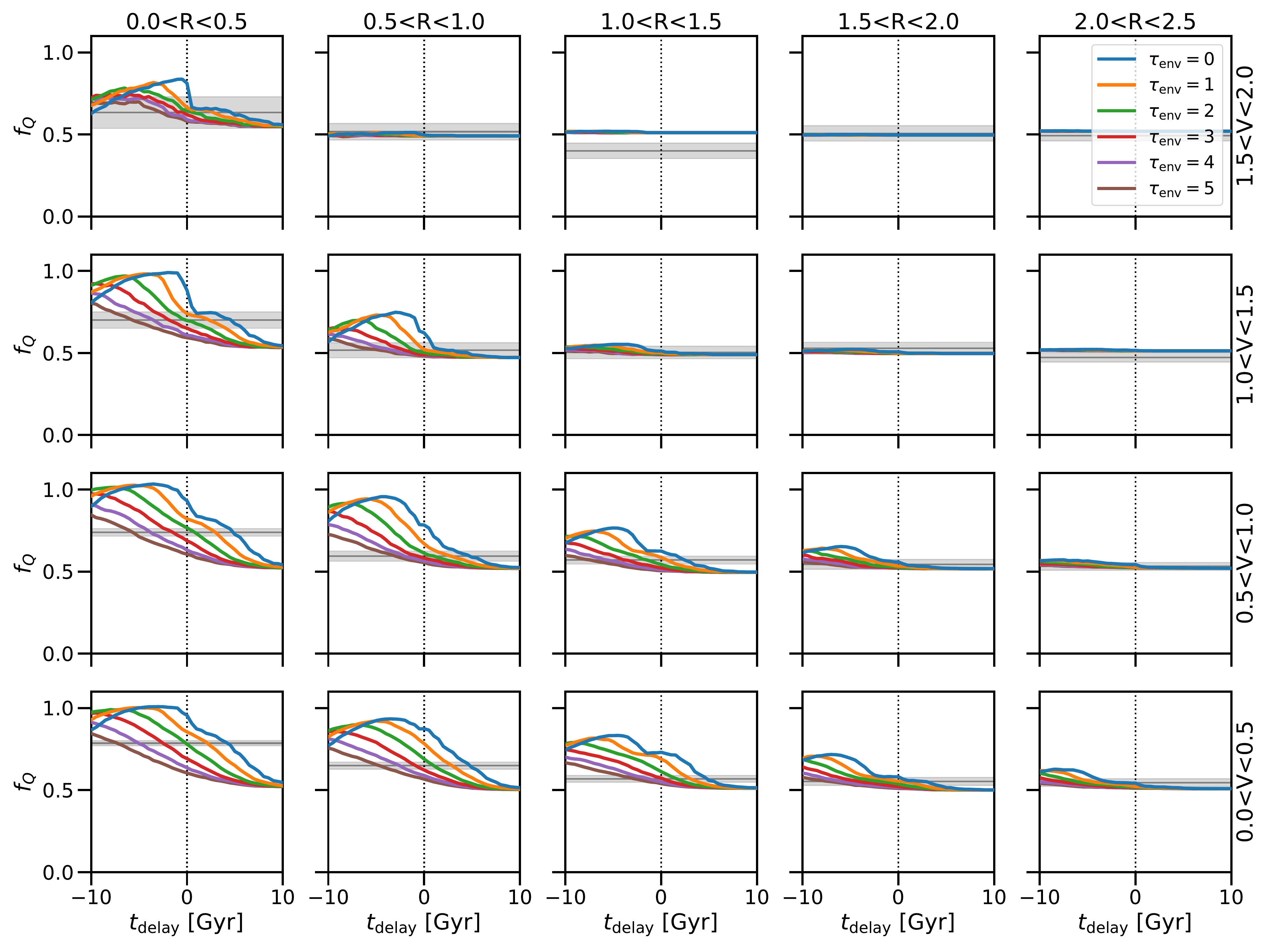}
	\includegraphics[width=1.7\columnwidth]{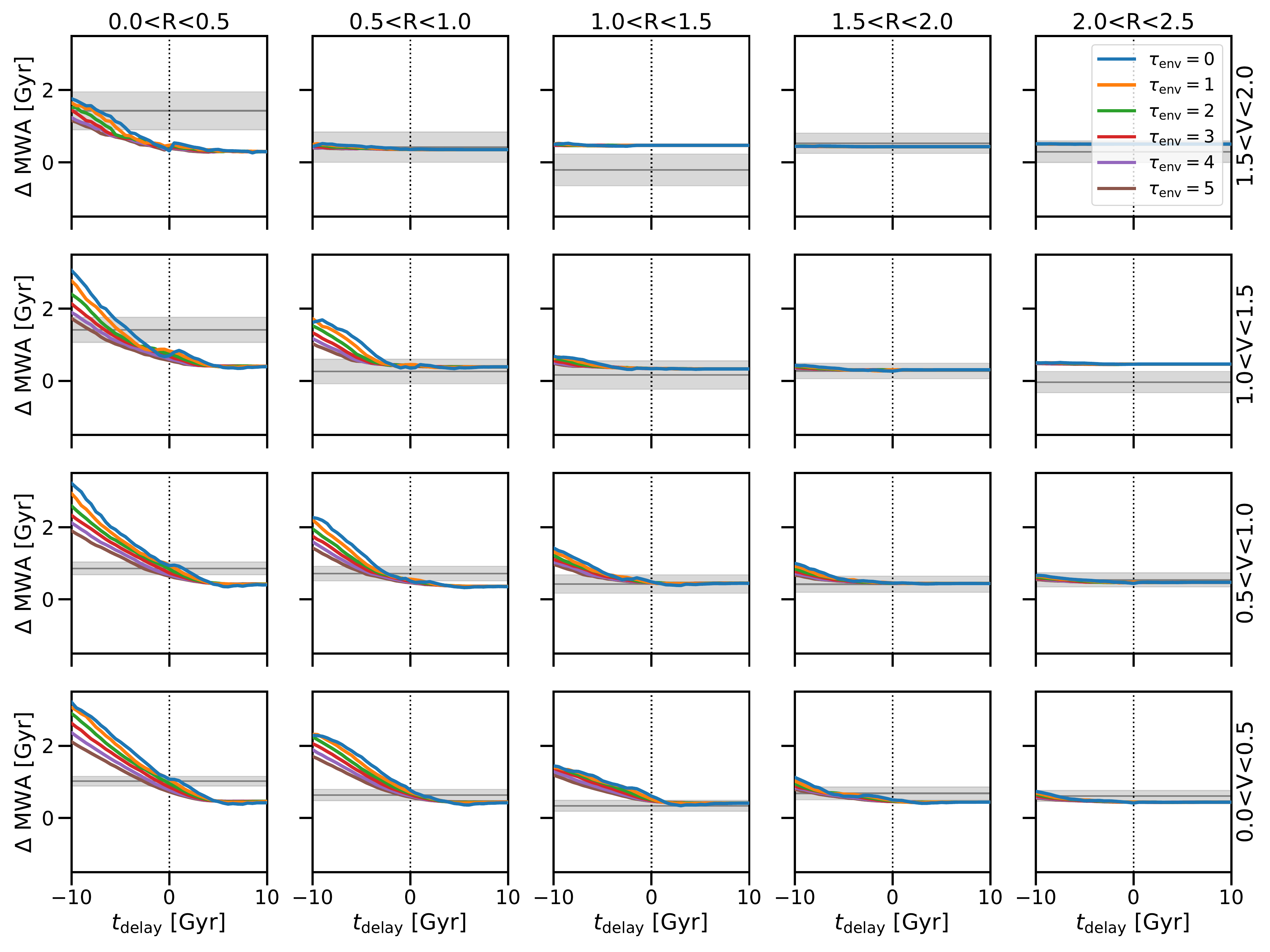}
    \caption{
    $f_\mathrm{Q}$ (upper multi-paneled figure) and mean $\Delta$MWA (lower figure) predictions for $10<\logMstellar<10.5$ galaxies for a range of models where galaxies quench after some delay time, relative to time of first pericentre. We show models run for a range of exponential suppression timescales. The SDSS mean values are shown with grey lines, with the shaded region showing the bootstrapped (over clusters) error on the mean. Analogous to Figs.~\ref{fig:fQ_tdelay_and_tau_model_lowMstellar} and \ref{fig:deltaMWA_tdelay_and_tau_model_lowMstellar}.
    }
    \label{fig:fQ_tdelay_and_tau_model_midMstellar}
\end{figure*}

Analogous to Fig.~\ref{fig:fQ_tdelay_and_tau_model_lowMstellar} and Fig.~\ref{fig:deltaMWA_tdelay_and_tau_model_lowMstellar}, we show our detailed model predictions for $10<\logMstellar<10.5$ galaxies in bins of PPS for $f\sbr{Q}$ and $\Delta$MWA in Fig.~\ref{fig:fQ_tdelay_and_tau_model_midMstellar}. Similar to our lower stellar mass bin, a somewhat different range of $\tdelay$ and $\tauenv$ values are preferred by $f\sbr{Q}$ and $\Delta$MWA. The joint best-fit for these parameters using both observables is given in Table~\ref{tab:best_fit_parameters}, with the contours shown in Fig.~\ref{fig:timescales_bestfits_fQplusMWA_break_the_degeneracy}.

\bsp	
\label{lastpage}
\end{document}